\documentclass[Review,sagev,times]{sagej}

\usepackage{breqn}

\usepackage{moreverb,url}

\usepackage[colorlinks,bookmarksopen,bookmarksnumbered,citecolor=red,urlcolor=red]{hyperref}

\usepackage[colorinlistoftodos,prependcaption]{todonotes}
\usepackage[normalem]{ulem}

\newcommand{\mbw}{{\mathbf{w}}}

\definecolor{red}{rgb}{0.00,0.00,0.00}
\newcommand{\cred}[1] {\textcolor{red}{#1}}

\newcommand\BibTeX{{\rmfamily B\kern-.05em \textsc{i\kern-.025em b}\kern-.08em
T\kern-.1667em\lower.7ex\hbox{E}\kern-.125emX}}

\def\volumeyear{2016}

\begin{document}

\runninghead{Salas}

\title{Improving the Estimation of the COVID-19   \cred{Effective} Reproduction Number using Nowcasting}

\author{Joaquín Salas\affilnum{1} }

\affiliation{\affilnum{1}CICATA Querétaro, Instituto Politécnico Nacional
}

\corrauth{Joaquin Salas,  Instituto Politécnico Nacional, Cerro Blanco 141, Colinas del Cimatario, Querétaro, México}

\email{salas@ieee.org}

\begin{abstract}
 As the interactions between people increases, the impending menace of COVID-19 outbreaks materialize, and there is an inclination to apply lockdowns. In this context, it is essential to have easy-to-use indicators for people to use as a reference. The  \cred{effective} reproduction number of confirmed positives, $R_t$, fulfill such a role. This document proposes a data-driven approach to nowcast $R_t$ based on previous observations' statistical behavior. As more information arrives, the method naturally becomes more precise about the final count of confirmed positives. Our method's strength is that it is based on the self-reported onset of symptoms, in contrast to other methods that use the daily report's count to infer this quantity. We show that our approach may be the foundation for determining useful epidemy tracking indicators.
\end{abstract}


\maketitle

\section{Introduction}

After a period of confinement due to the presence of COVID-19 and facing economic and social pressures, societies start to open up, seeking to return to productive, sport, and recreational activities. As the interactions between people increase, the impending menace of outbreaks materializes. Naturally, there is a tendency to apply once again lockdowns, in what has been called {\it  the hammer and the dance}~\cite{pueyo2020coronavirus}. In this context, it is essential to have easy to apply indicators for people to use as a reference. 
\cred{
At the beginning of the infection, when all population members are susceptible, the average number of illnesses that an infected person originates is called the basic reproduction number, $ R_0 $. Sometime after the beginning of the infection, and with considerably more practical utility, one may want to know the effective reproduction number, $R_t$~\cite{ma2020estimating}. }
When $R_t$ is higher than one, the number of infected people grows exponentially, {\it i.e.}, their number will double in a short period. When $R_t$ is less than one, the epidemic will tend to disappear. However, estimating $R_t$ accurately at the required level of geospatial resolution is a complex problem.

Although applicable to any country, let us take the case of Mexico as an example. The records generated by the epidemiological surveillance system contain information that includes, among other predictors, the number of confirmed positives, deaths, and suspects. Daily, the Ministry of Health informs the public about the status of its records~\cite{dataset}. However, the data it discloses updates records of events that occurred in the past, sometimes as far as  \cred{90} days ago. At other times, with a significant frequency, the records that were previously released are discarded. Although publishers often drop these erroneous entries overnight, there have been cases of records eliminated after more than 50 days.

Besides the integrity of the information, there are other difficulties in tracking the epidemy inherent to the pandemic and interesting for researchers, decision-makers, and the general public. SARS-CoV-2 is an airborne virus~\cite{bahl2020airborne}, which infects some people without causing symptoms~\cite{nishiura2020estimation}. On a significant number of occasions, people begin to spread COVID-19 before they start to feel sick~\cite{li2020substantial}. Also, each infected person reacts differently and will have, if anything, a different latency and incubation period~\cite{he2020temporal}. People will have a different contagious period, manifested with inequal intensity during that time~\cite{byrne2020inferred}. Although the symptoms are known, one may reveal them  differently. People will require different types of medical attention, which may or may not require hospitalization~\cite{garg2020hospitalization}. In some cases, someone ill may need or not a ventilator~\cite{covid2020forecasting}. Eventually, a given person may recover, possibly with sequels, or will pass away~\cite{salas2020data}. About the whole process, we begin to have some statistical knowledge on which we can develop models. In this paper, propose a data-driven approach that leverage experience to create a simple, yet effective nowcasting method for $R_t$ that can be used by policy-makers as well by the general public. Our main contribution is an approach to use past observations to generate plausible sequences of estimates for the number of confirmed positive cases that could have possibly occurred in the recent days to compute variations of the \cred{effective}  reproduction number.

We base our method on the statistical behavior of previous observations. As more information arrives, the estimation naturally becomes more precise about the final count of confirmed positives.
In the next section, we review the literature about related methods.
 \cred{ Then, we proceed to discuss the intrinsic delay in information flow that exists in the process of detecting a COVID-19 confirmed positive and detail our approach to estimate plausible sequences for the number of infected people. This insight leads us to review the underlying method we employ to calculate the effective reproduction number using the health reports available. After showing some results of our implementation of the nowcasting method for $R_t$, we conclude our study by discussing and delimiting our findings and delineating some potential research lines.}

\section{Related Literature}
Though recent, COVID-19 has kickstarted some novel ideas to track it reliably.
The research effort to nowcast the basic reproduction number can be classified in either mechanistic approaches, Bayesian approaches, or a hybrid combination of both.

\subsection{Mechanistic Approaches}
Wang {\it et al.}~\cite{wang2020spatiotemporal} developed a hybrid model to complement the dynamics of the SIR (\cred{Susceptible} , Infectious, Recovered) model with spatiotemporal analysis.
The space-time component is modeled, at the start, with a Poisson distribution to describe rare events. Then, they complemented it with a negative binomial random model during over-dispersion.
Balabdaoui and Mohr~\cite{balabdaoui2020age} propose an age-stratified discrete compartment model as an alternative to SIR type models. Their approach follows the trajectory of individuals that includes the exposed, the asymptomatic, the symptomatic infectious, the symptomatic in self-isolation, the patients in the intermediate care unit, and the patients in the intensive care unit.
Masjedi {\it et al.}~\cite{masjedi2020nowcasting} compares  phenomenologic  and mechanistic models. The former based on generalized Richards models~\cite{richards1959flexible} (an extension of sigmoid functions) and the latter on a modified
SEIR (Susceptible, Exposed, Infectious, and Recovered)  model. They fit the models with observed data to forecast the next month.
They observe that although phenomenologic  models fit the data, they are not reliable for decision-making. In contrast, SEIR models predicted the phenomena better. Contaldi~\cite{contaldi2020covid} presents SIRFH, an extension of the SIR  model that tracks hospitalizations and hospital-based fatalities introducing additional differential equations. The estimation for the basic reproduction number derives from the solution to this extended model. Finally, Annan and Hargreaves~\cite{annan2020model} produce a nowcasting method based on the SEIR model.  To calibrate the parameters, they use observational data and a Bayesian approach. Annan and Hargreaves' analysis includes the uncertainties associated with deaths' stochastic nature, the reporting errors, and the model itself.

\subsection{Bayesian Approaches}
Altmejd {\it et al.}~\cite{altmejd2020nowcasting} present a model based on the removal method~\cite{pollock1991review}, where one extracts batches of a fixed population. Their models deal with lags arising from the calendar patterns, where events reported during the weekends are less. Their Bayesian approach uses a likelihood that considers the number of reports by day of the week, and priors with improper uniform distribution. Their model provides better estimates than seven days averages.  Schneble {\it et al.}~\cite{schneble2020nowcasting} present a nowcasting model based on the number of deaths, as quantifying their correct number is more reliable than for infected people. Their epidemic spread model considers region and age-specific Poisson distributions, where they consider lag to report. They model the effect of age, gender, weekday, and location as a quasi Poisson distribution. Then, they infer a posterior using a Gaussian prior. For nowcasting, they model the delay as a random variable which will provide death counts. They distribute these death counts as a quasi-binomial distribution. Chitwood {\it et al.}~\cite{chitwood2020bayesian} propose to use a Bayesian framework for nowcasting. They take into account delayed and incomplete reporting.
They assume that one can understand the COVID-19 complex spread system by examining the individual components. In that model, they consider the uncertainty that results from available diagnosis and delays in the estimation of disease progression and reporting systems. Lastly, Abbot {\it et al.}~\cite{abbott2020estimating} employ a quasipoisson regression model to estimate the spread rate. Interestingly, they base their analysis on the reported dates for the confirmed positives and infer the symptom onset through statistical modeling.

\begin{figure*}
    \centering
    \begin{tabular}{cc}
    \includegraphics[width=2.25in]{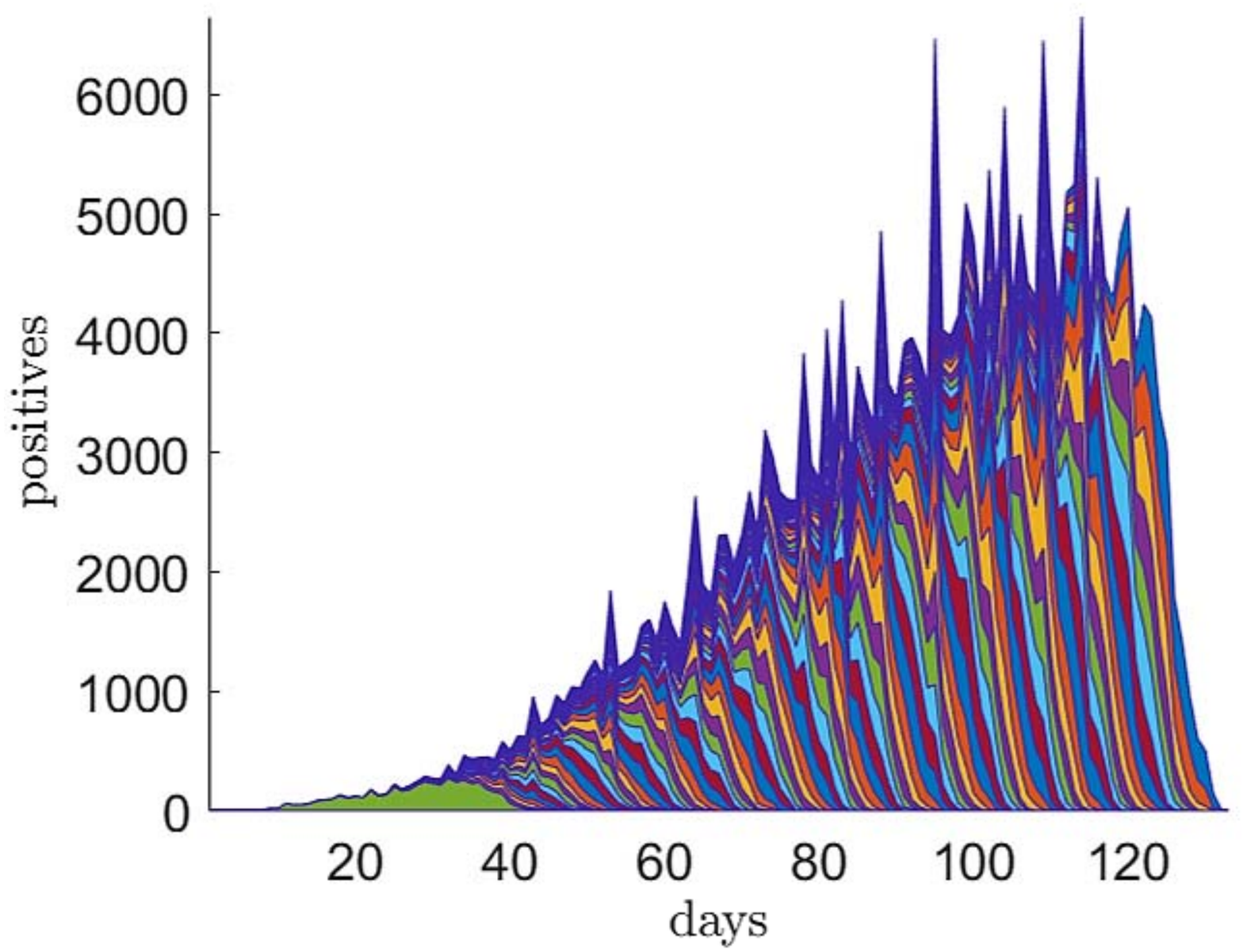} &
    \includegraphics[width=2.25in]{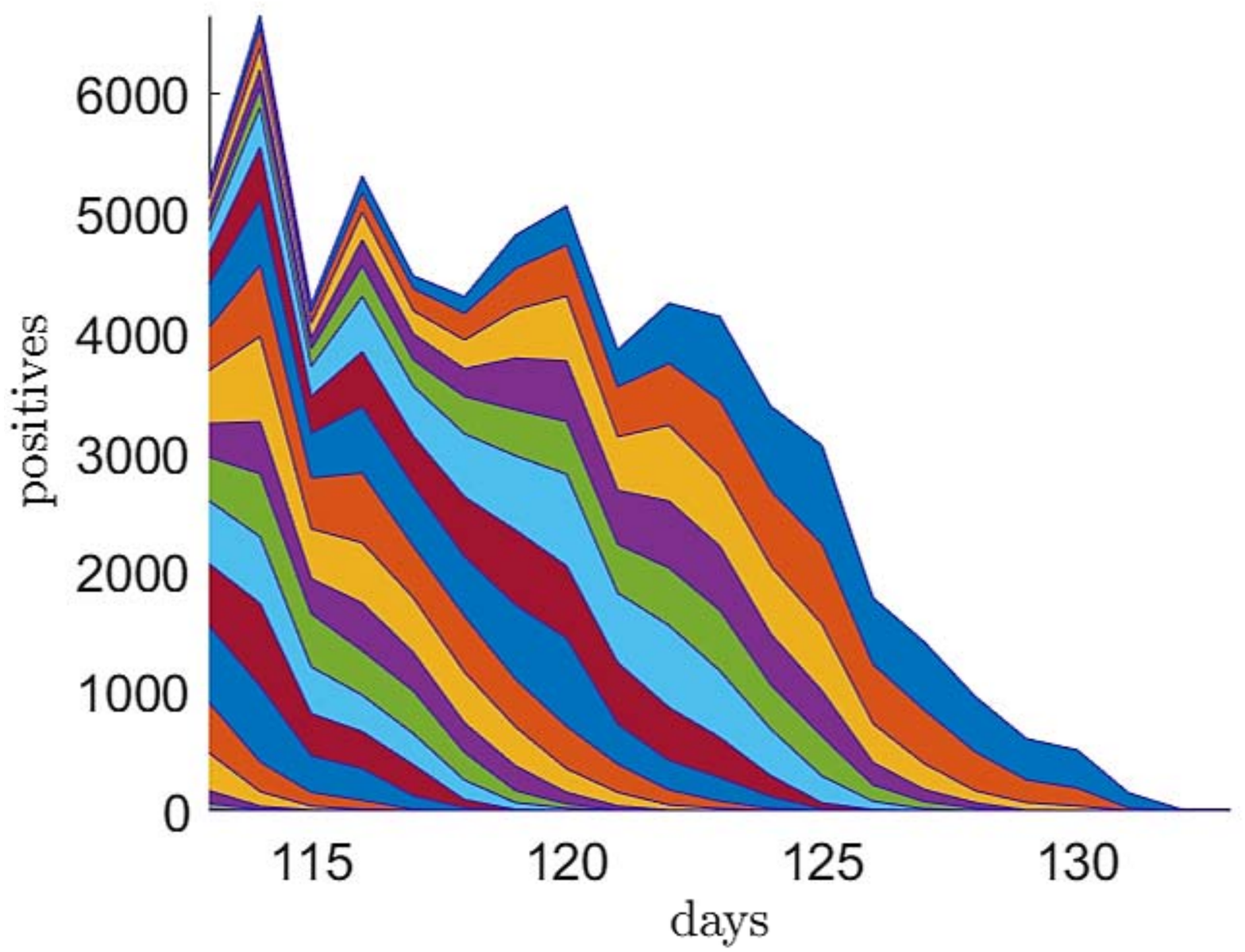}\\
    (a) & (b) \\

    \end{tabular}
    \caption{Delays in reporting. The \cred{horizontal axes show}  the day of onset. The vertical \cred{axes indicate}  the number of patients confirmed positives. \cred{In (b), we zoom in on the last 30 days shown in (a). }Each layer is an update to registers in the past.}
    \label{fig:delay}
\end{figure*}


\section{Characterizing the Update Pattern}
\label{sec:delays}
In our approach, we characterize the frequency at which the counting updates of COVID-19 confirmed positives occur. In this section, we analyze the origin of such delays and describe the form we model them.

\subsection{Delays in the Report of Confirmed Positives}

Declaring a person confirmed positive involves a complex process that may take days, even nowadays, when it is of paramount importance to achieve certainty for decision-making. Just consider the case of a person showing symptoms related to COVID-19~\cite{menni2020real} that decides to visit the physician. After an interview to collect some necessary clinic information, the physician chooses to take either a sample from the nasopharynx using a long swab~\cite{petruzzi2020covid} or a CT (Computer Tomography)~\cite{long2020diagnosis}. In some places, the sample can be analyzed via the RT-PCR(reverse-transcription polymerase chain reaction)~\cite{yang2020patients} {\it in situ} with results on the same day but frequently it may take a week or longer to be processed. Afterward, the results will be uploaded in computer systems and summarized for analysis.

\begin{figure*}
    \centering

    \begin{tabular}{ccc}
    \includegraphics[width=1.5in]{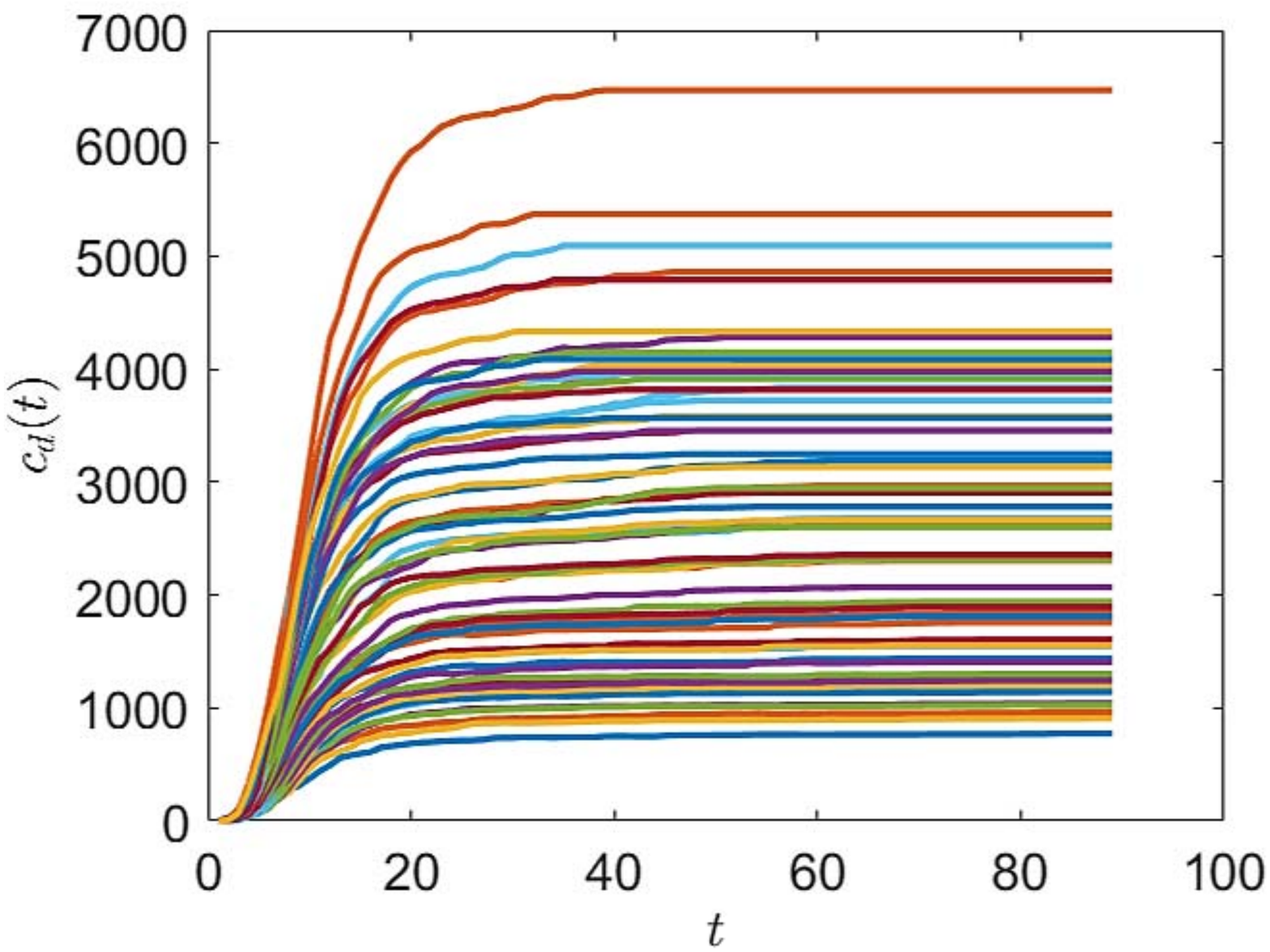} &  \includegraphics[width=1.5in]{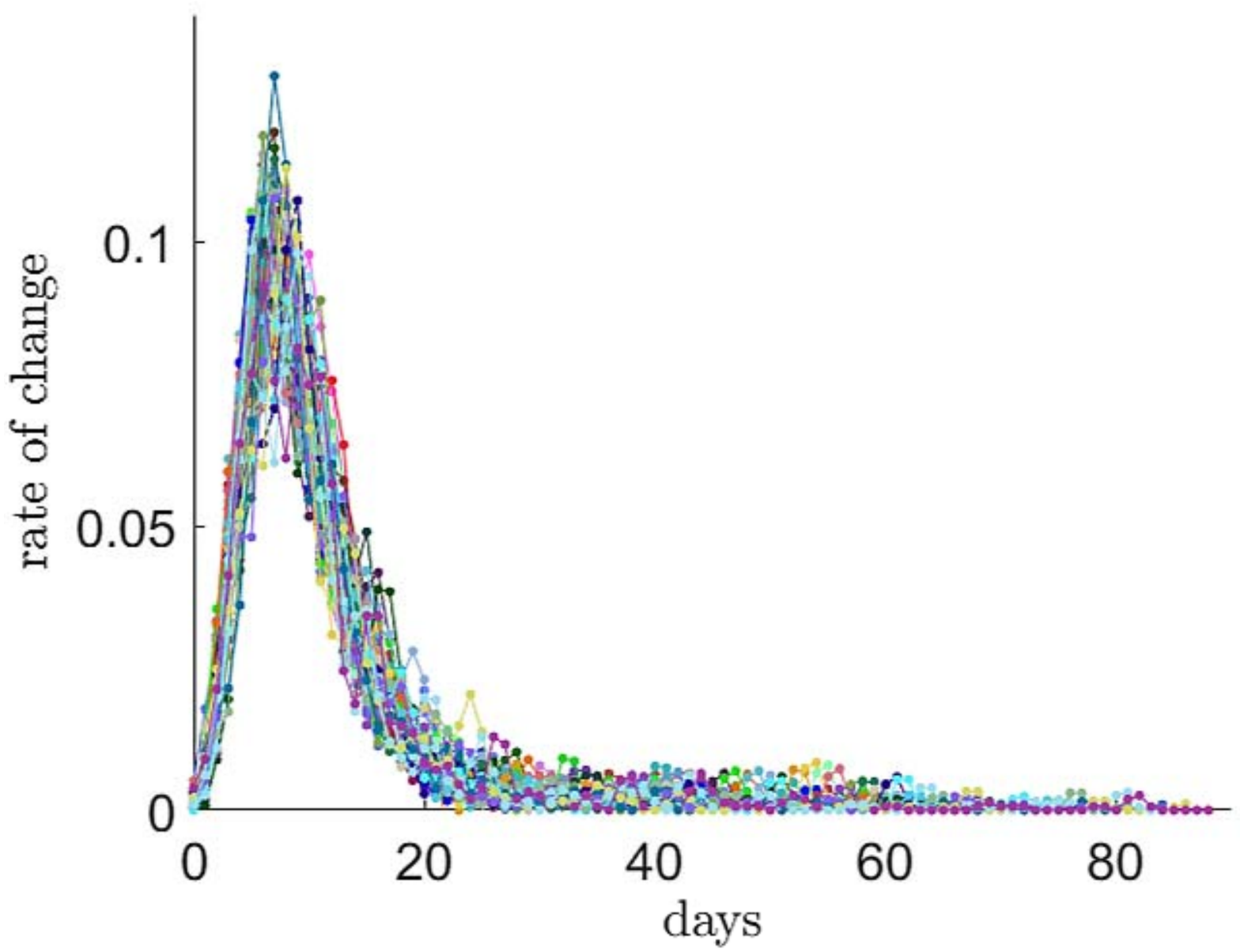} &
    \includegraphics[width=1.5in]{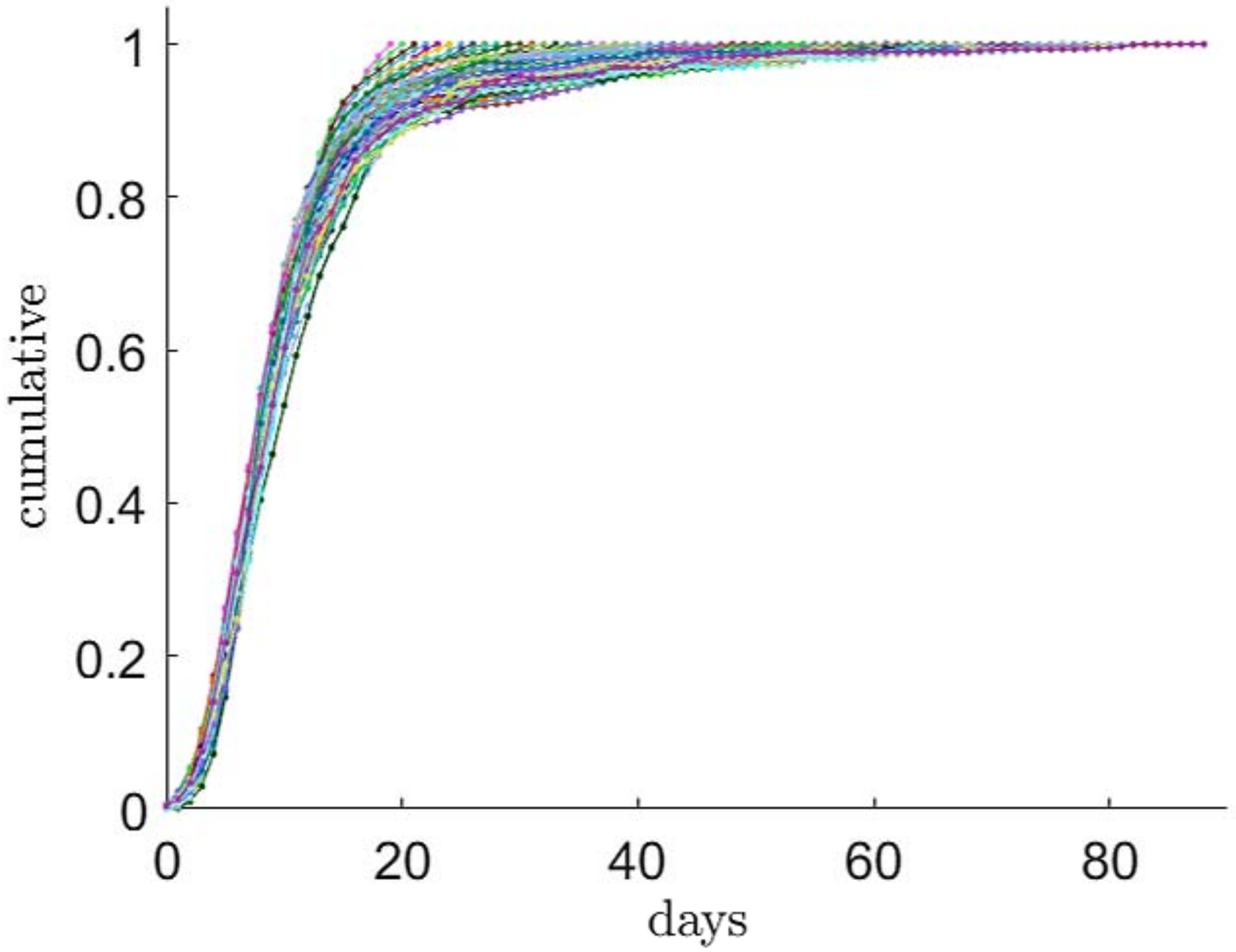}\\
    \cred{(a)} & \cred{(b)} & \cred{(c)}\\
    \end{tabular}
    \caption{Confirmed positives. (a) As the days pass, updates eventually level off to a final count for a given day $C_t(D)$. (b) and (c) show the normalized daily number of reported confirmed positive and the accumulated number of cases. Our method relies on the assumption that it is possible to model the daily variations with a data distribution.}
    \label{fig:confirmed-positives}
\end{figure*}

In Figure~\ref{fig:delay}, we illustrate the effect of delays in reporting using the data set made public by the Mexican Health Ministery~\cite{dataset}. The horizontal and vertical axes show the day of onset and the number of confirmed positive cases. Each layer corresponds to the number of cases added to a prior date.  Although the number of updates may be significant for a given day, they eventually converge to the total number of confirmed positives for that day, $C_t(D)$, for $D$ large, and where  $t$ expresses the day of interest (see  Figure~\ref{fig:confirmed-positives}(a)). If we divide the daily accumulated of confirmed positives $C_t(\delta)$, for a given day  $\delta$, by  $C_t(D)$, the cumulated distribution will tend to one. We illustrate this in  Figure~\ref{fig:confirmed-positives}(b)-(c), where we show both, the rate of daily change and the cumulative change. Our approach aims to characterize the variations we observe in these distributions to develop a model for nowcasting.

\subsection{Compounded Rate of Change}
\label{sec:compound}

We aim to estimate the number of confirmed positive cases $C_t(D)$ for the day $t$  using the following   $\delta$ days of reports available. In principle,  we would learn about $C_t(D)$  when $D$ is cosiderably large. But in practice, $D$ can be as short as one month and a half of daily updates. Given the number of confirmed positives $\delta$ days after day $t$, $C_t(\delta)$, the number of confirmed positives on day $C_t(\delta+1)$  can be expressed as
\begin{equation}
    C_t(\delta+1) = C_t(\delta) (1 + \rho_t(\delta)),
    \label{eq:recurrence}
\end{equation}
where $\rho_t(\delta)$ is the rate of change from one day $\delta$ to the next  $\delta + 1$, for reference  day $t$. If we solve the recursion, we will have the  expression
\begin{equation}
    C_t(D) = C_t(0)\prod_{\delta=0}^{D-1} (1 + \rho_t(\delta)),
    \label{eq:confirmed-positives-full}
\end{equation}
where one assumes that the daily rate changes over time.
In the cases we are studying, the curves expressing the rate of change of the number of confirmed positive relative to the day before, for a different starting day, seem to be somewhat consistent over the samples.   \cred{We} model $\rho_t$ as a random variable, \cred{which follows a probability distribution we may infer from } the experimental samples. Then,  on the day  $t+\delta$ , the best-guess prediction for the number of confirmed positive, $C_t(\delta)$,  is
\begin{equation}
C_t(D) = C_t(\delta) (1 + \rho_t^D(\delta)),
\label{eq:confirmed-positives-short}
\end{equation}
where our newly defined random variable $\rho_t^D(\delta)$ expresses the rate of change from day $\delta + t$ to day $D$.
In our approach, we model $\rho_t^D(\delta)$ as a random variable with different  \cred{distribution} for each day $\delta$, for more fine-grained or longer-term prediction.
One may find the relationship between $\rho_t^D(\delta)$ and $\rho_t(\delta)$ by noting that (\ref{eq:confirmed-positives-full}\cred{)} and (\ref{eq:confirmed-positives-short}) solve for   \cred{$C_t(D)$} as
\begin{equation}
 C_t(0)\prod_{\delta=0}^{D-1} (1 + \rho_t(\delta))  =  C_t(\delta) (1 + \rho_t^D(\delta)).
\end{equation}
Expanding $C_t(\delta)$ using the recurrence relationship in (\ref{eq:recurrence}), we have
\begin{equation}
 C_t(0)\prod_{\delta=0}^{D-1} (1 + \rho_t(\delta))  = C_t(0)\prod_{d=0}^{\delta-1} (1 + \rho_t(d)) (1 + \rho_t^D(\delta)),
\end{equation}
from where, after eliminating for the common factors, solving for $\rho_t^D(\delta)$ results in
\begin{equation}
\rho_t^D(\delta) =  \prod_{d=\delta}^{D-1} (1 + \rho_t(d)) - 1.
\end{equation}

\section{\cred{Effective}  Reproduction Number $R_t$}
\label{sec:basic}
Given a particular sequence of the observed number of infected people $\{C_0(\delta), C_1(\delta-1) \dots, C_t(0)\}$, and the argument of the number of days the report has been updated, we aim to nowcast the basic reproduction number $R_t$, {\it i.e.},  given the distribution of the rate of change $\rho_t^D(\delta)$, we generate ensembles of sequences aiming to estimate $\{C_0(D), C_1(D) \dots, C_t(D)\}$  before proceeding to calculate $R_t$. We first review {\it EpiEstim}, a  method proposed by Cori {\it et al.}~\cite{cori2013new}, to estimate $R_t$ from the observed number of cases.

Cori {\it et al.}~\cite{cori2013new} proposed a Bayesian framework to compute $R_t$, where the number of infected people observed at day $t$, $C_t$, follows a Poisson process. In a simplification, they assume that the daily observations of infected people are independent. Thus, one may express the  likelihood of observing a sequence of infected people between day $t-\tau -1$ and day $t$ as~\cite{cori2013new}
\begin{dmath}
P(C_{t -\tau +1}, \dots,C_t\mid C_0, \dots, C_{t-1}, \mbw, R_{t,\tau}) =
\prod_{s = t - \tau +1}^t \frac{(R_{t,\tau} \Lambda_s)^{C_s} e^{-R_{t,\tau} \Lambda_s}}{C_t!},
\end{dmath}
where the transmisibility $R_{t,\tau}$ is assumed to be constant over the period $[t-\tau+1, t]$,  $\Lambda_t = \sum_{s=1}^{t} C_{t-s} w_s$ is the total infectiousness of infected people at time $t$, and
 $\mbw = (w_1, \dots, w_t)^T$ is a mass density probability profile of infectivity profile  for an individual. Cori {\it et al.}~\cite{cori2013new} assume that
 the  \cred{effective} reproduction number  $R_{t, \tau}$ is a random variable which probability  follows a  Gamma distribution as~\cite{cori2013new}
\begin{equation}
P(R_{t, \tau}) = \frac{R_{t,\tau}^{a-1}}{\Gamma{(a)} b^a}  e^{-R_{t,\tau}/b},
\end{equation}
where $a$ and $b$ are the parameters of shape and scale. Since the Poisson and Gamma probability distributions are conjugate, one can express the posterior in closed form, again as a Gamma distribution, as~\cite{cori2013new}
\begin{dmath}
P(C_{t - \tau +1}, \dots, C_t, R_{t,\tau}\mid C_0, \dots, C_{t-\tau}, \mbw)
\propto R_{t,\tau}^{\alpha - 1}
    e^{R_{t,\tau}/ \beta}
\prod_{s = t - \tau + 1}^t \frac{\Lambda_s^{C_s}}{C_s!},
\end{dmath}
from where  the mean $\alpha$ and standard deviation $\beta$ are given by
\begin{equation}
\alpha = a + \displaystyle{\sum_{s=t - \tau +1}^t C_s}  \mbox { and } \beta = \frac{1}{\displaystyle{\sum_{s=t-\tau+1}^{t} \Lambda_s} +\frac{1}{b}}.
\end{equation}

\begin{figure*}[th]
    \centering
    \begin{tabular}{ccc}
     \includegraphics[width=1.5in]{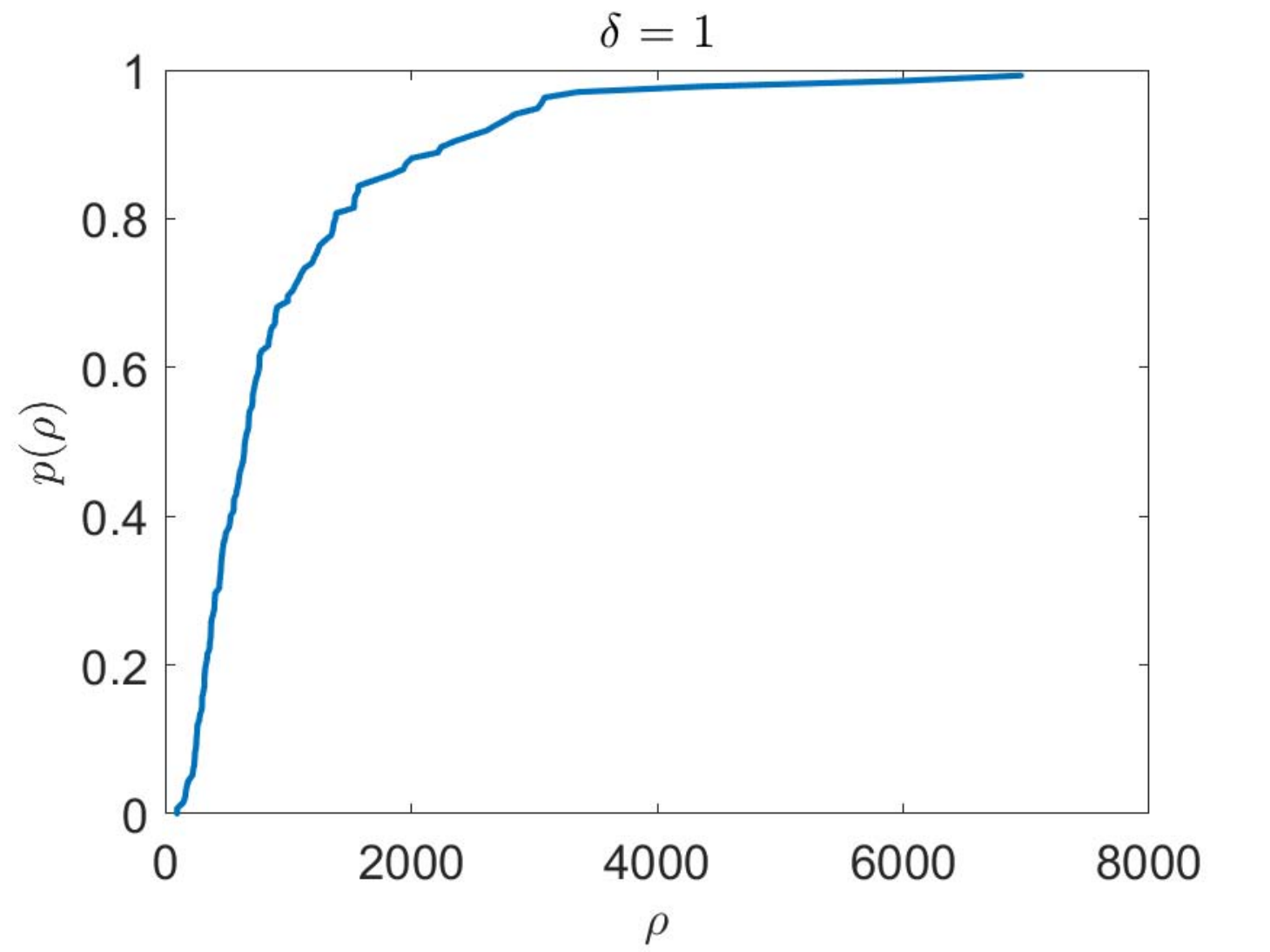} &
    \includegraphics[width=1.5in]{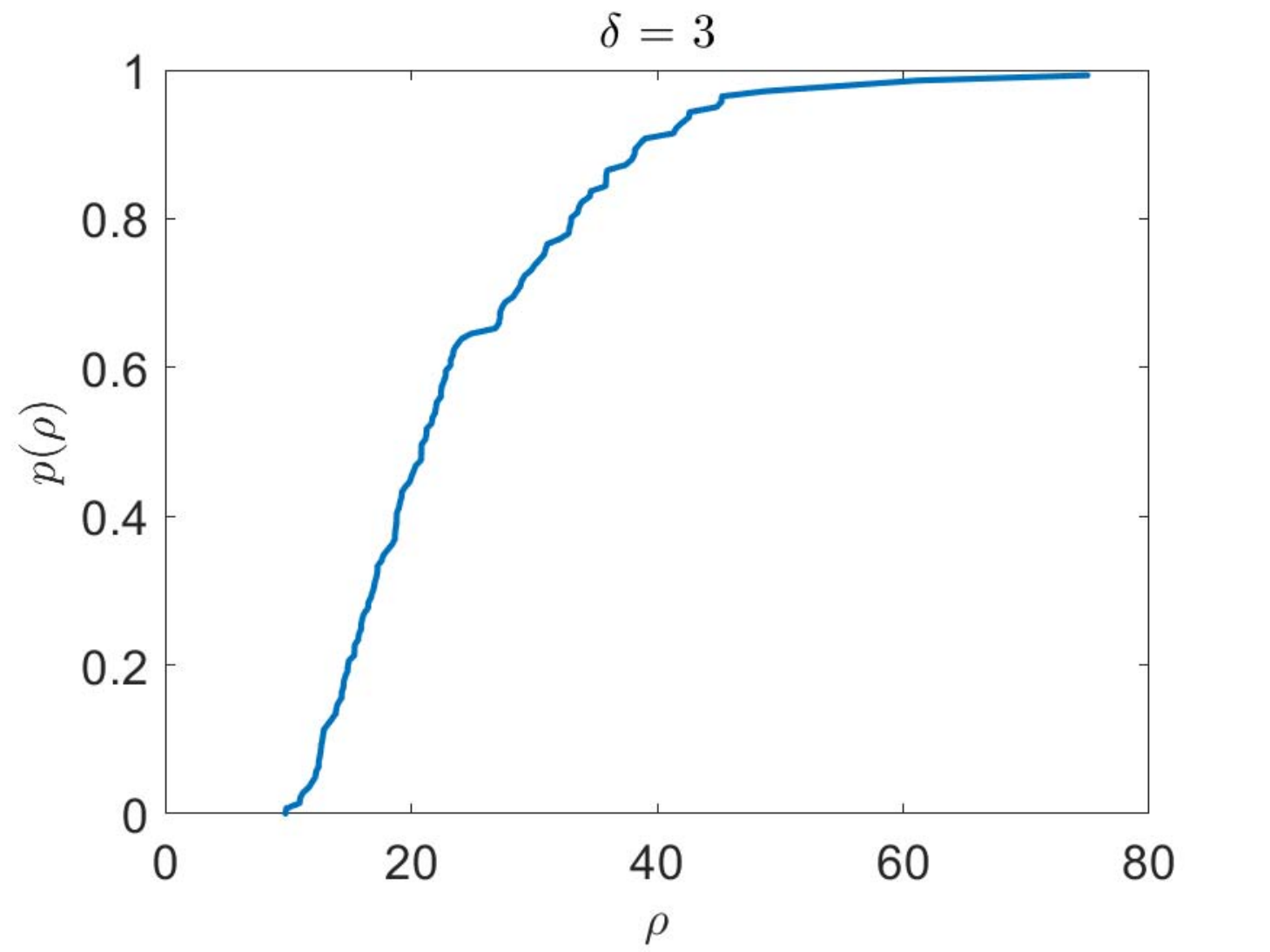} &
    \includegraphics[width=1.5in]{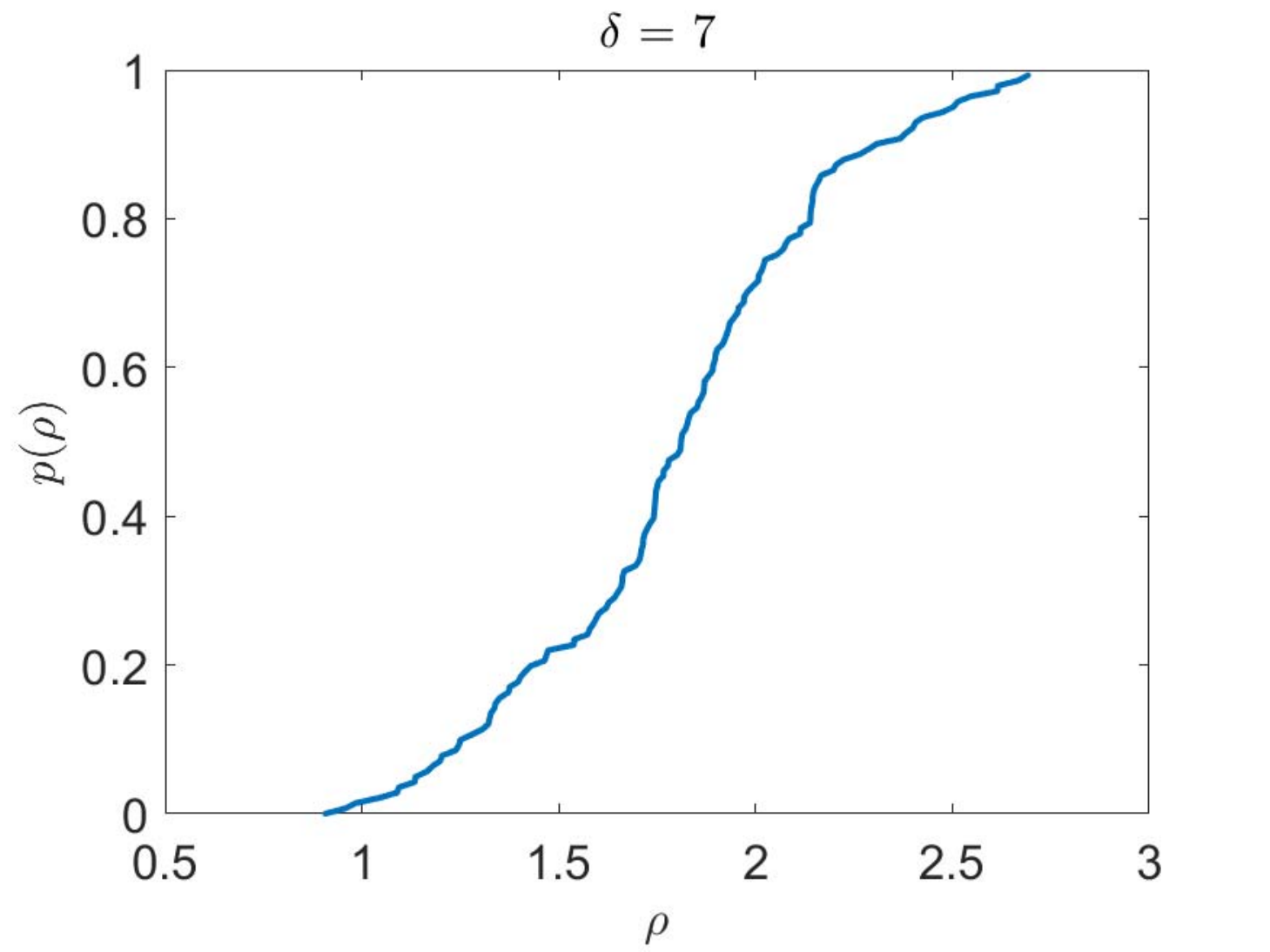}\\
    (a) & (b)  & (c)\\
    \includegraphics[width=1.5in]{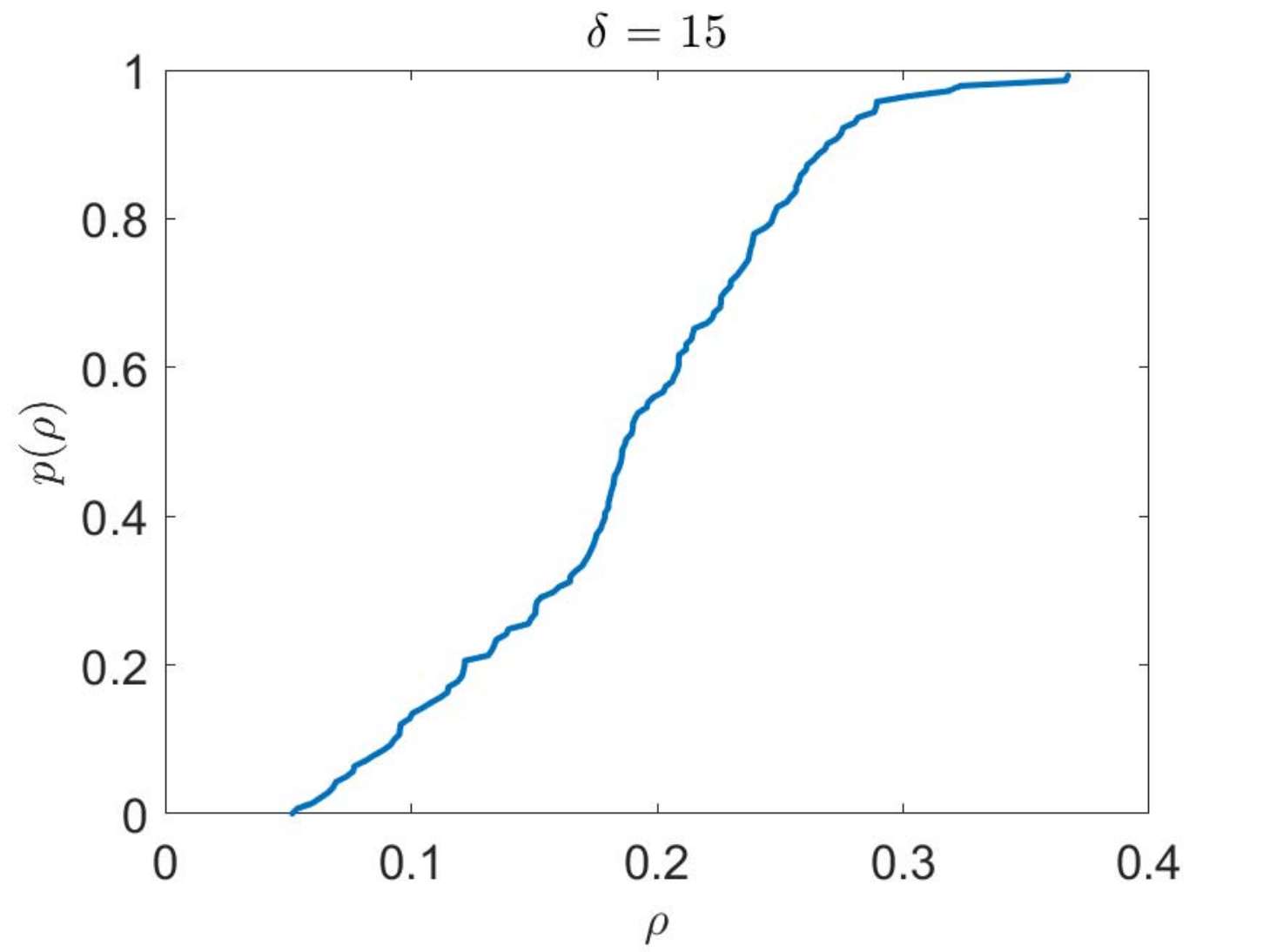} &
    \includegraphics[width=1.5in]{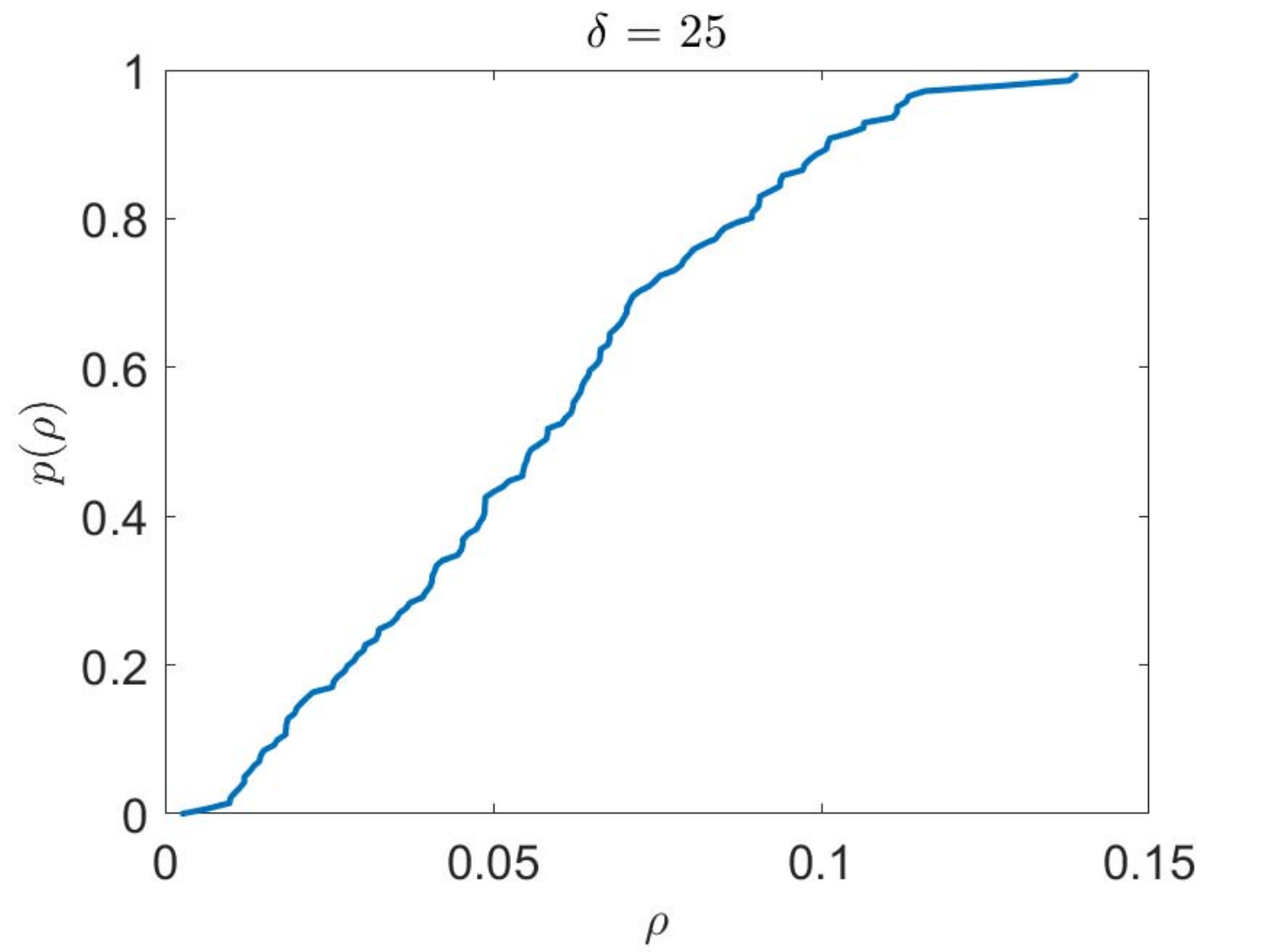} &
    \includegraphics[width=1.5in]{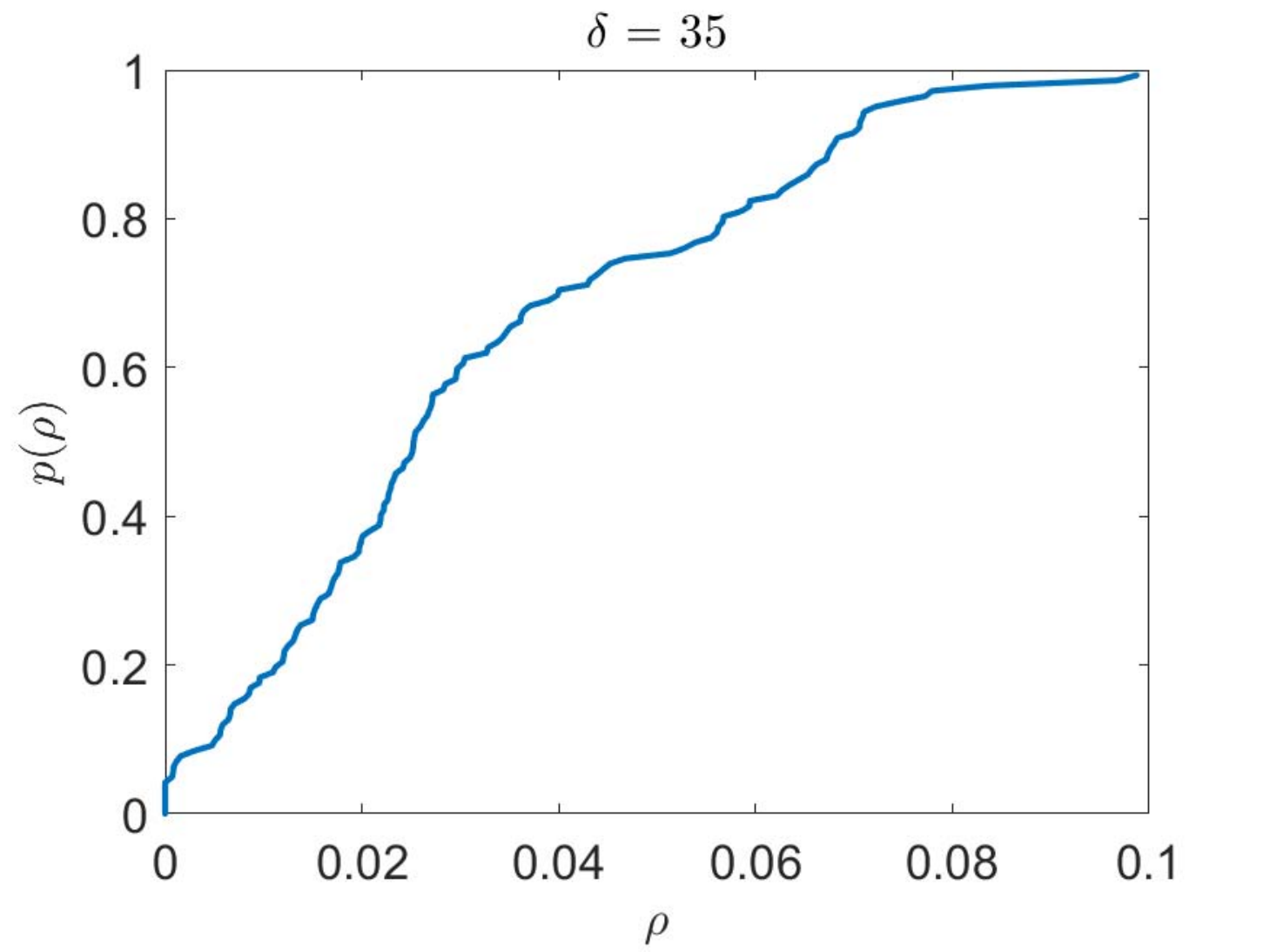}\\
    (d) & (e)  & (f)\\

    \end{tabular}
    \caption{\cred{Empirical Distribution Function}. \cred{Out of the historical observations, we construct distributions for $\rho$} Here, we show examples for $\delta=1, 3,7,15,25, 35$.}
    \label{fig:fit}
\end{figure*}

Given $C_t(\delta)$, the information about the number of infected people $\delta$ days after the day of interest $t$, and the model for the probability function for $\rho_t^D(\delta)$, we produce $N$ random samples which will correspond to the number of people infected that day. We then compute $R_t$ for each of the sequences using the model proposed by Cori {\it et al.}~\cite{cori2013new}. Finally, we calculate the mean and standard deviation for $R_t$ to provide the most likely value and uncertainty at one standard deviation. To take into account the difference between the  accepted values for the average incubation (five days)~\cite{he2020temporal} and latency periods (three days)~\cite{li2020substantial}, we represent them  two days before $t$.

\begin{figure*}[t]
    \centering
    \begin{tabular}{cc}
     \includegraphics[width=2.25in]{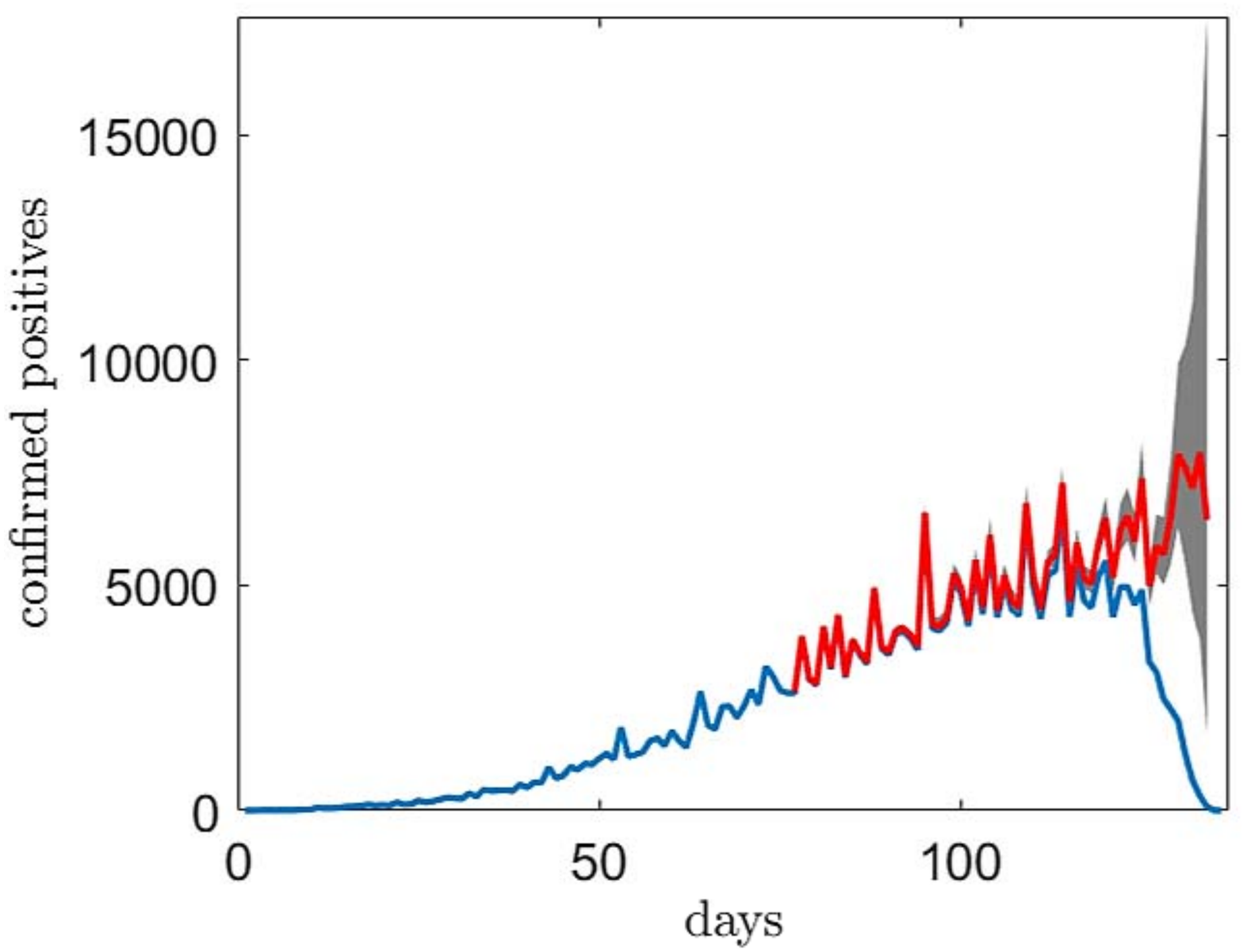} &
    \includegraphics[width=2.25in]{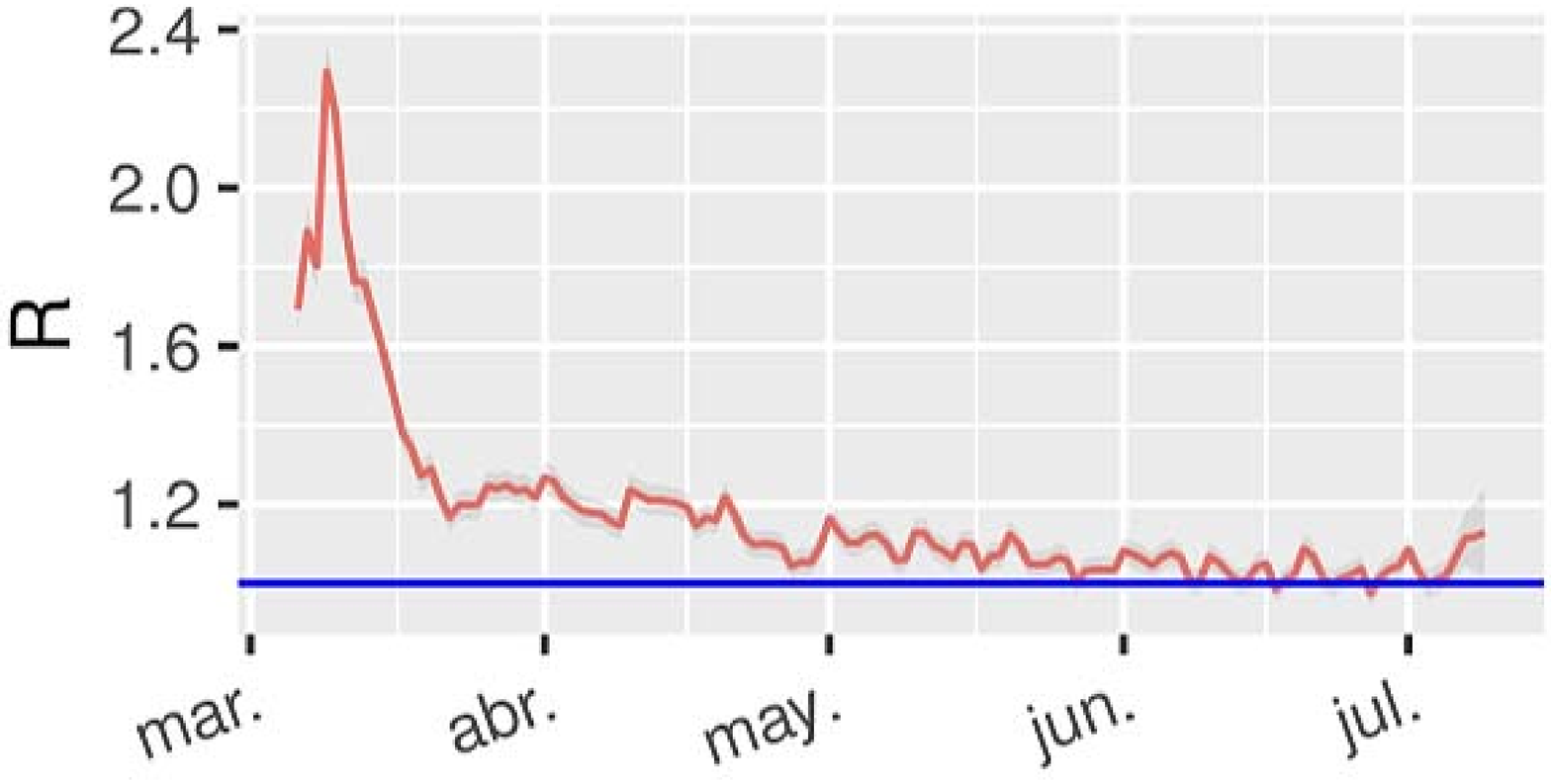}\\
    (a) Nowcasting the infected & (b) $R_t= 1.3\pm 0.113$\\
       \includegraphics[width=2.25in]{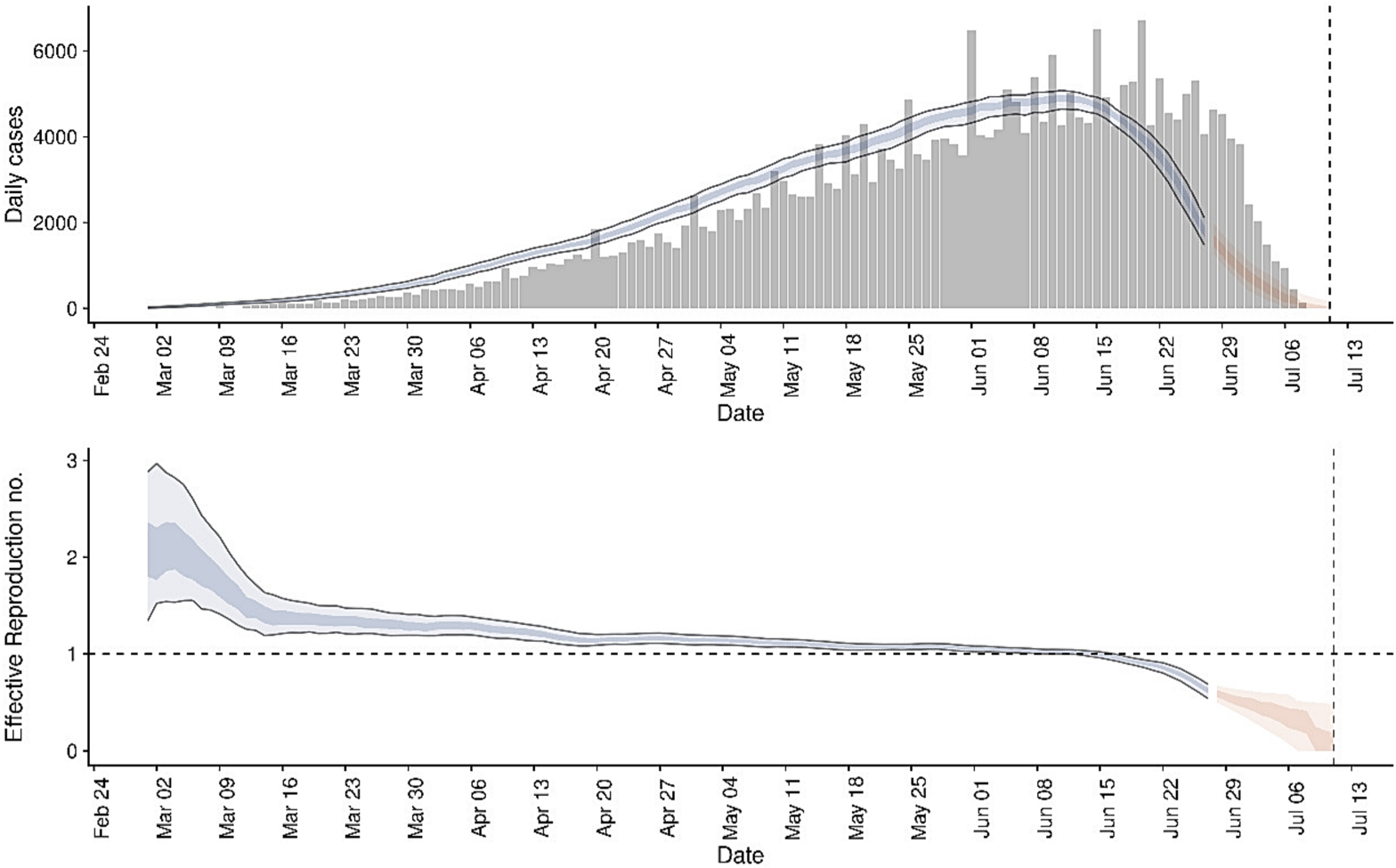} &
    \includegraphics[width=2.25in]{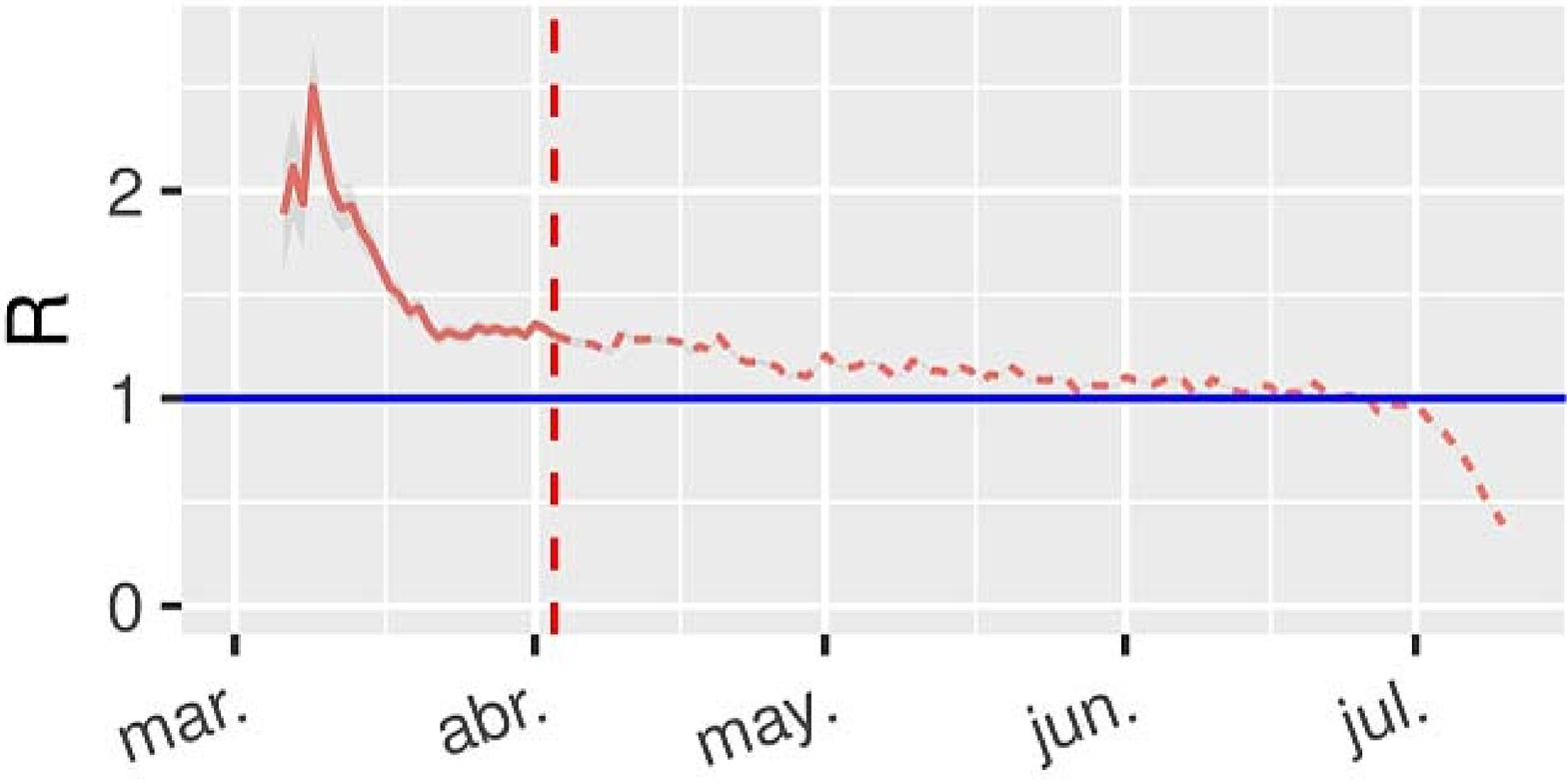}\\
    (c) & (d)   Without nowcasting \\

    \end{tabular}
    \caption{Qualitative comparison of the model's output. For the daily confirmed cases (a), we show Cori's output {\it et al.} algorithm to compute $R_t$. We fed the data to the method provided by Abbott {\it et al.}~\cite{abbott2020estimating}, which seems to follow the reported daily cases. Finally, we illustrate the output of our approach, including its area of uncertainty.  }
    \label{fig:comparison}
\end{figure*}

\begin{figure*}
    \centering
    \begin{tabular}{cccc}
     \includegraphics[width=1.125in]{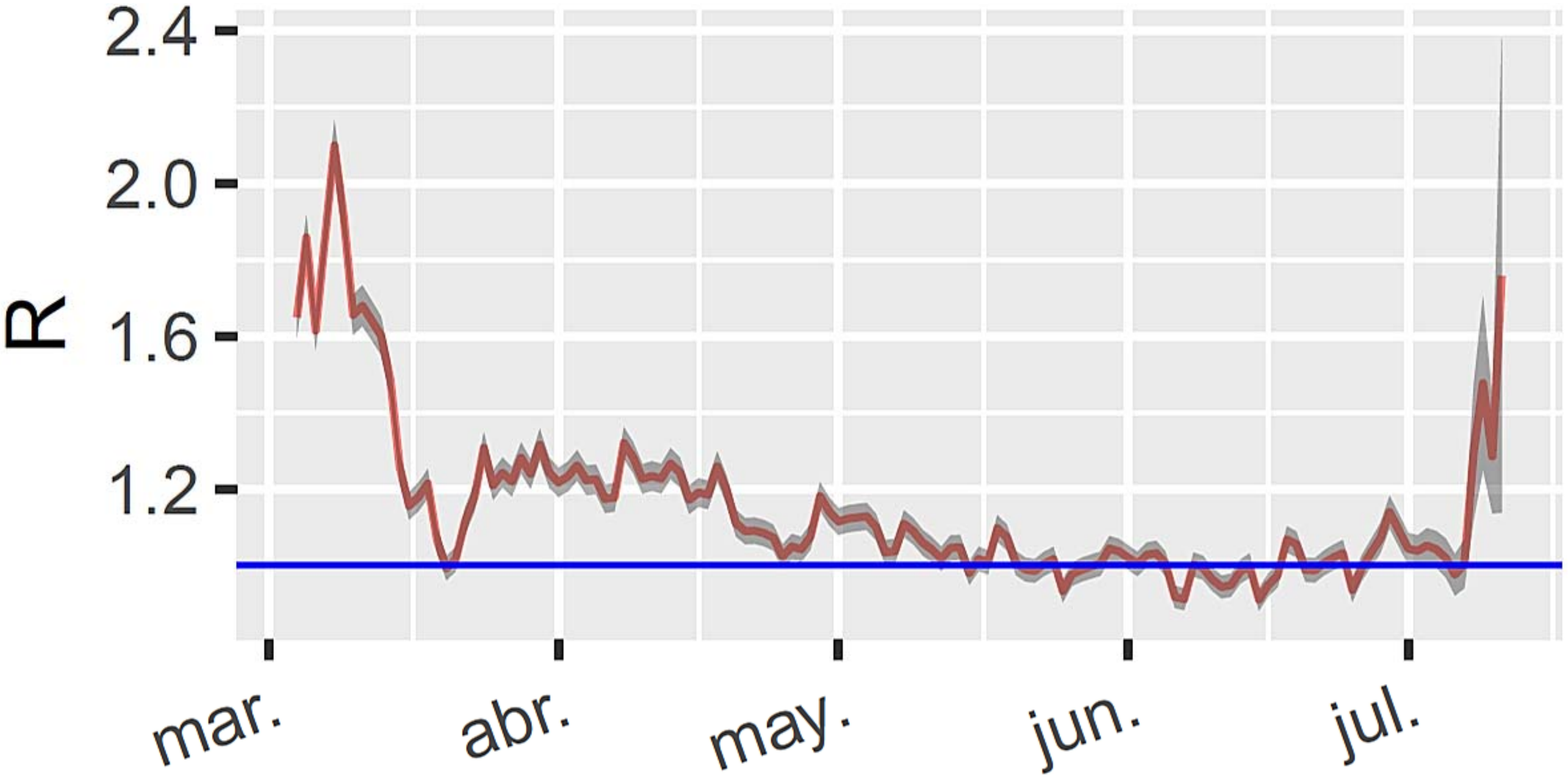} &
    \includegraphics[width=1.125in]{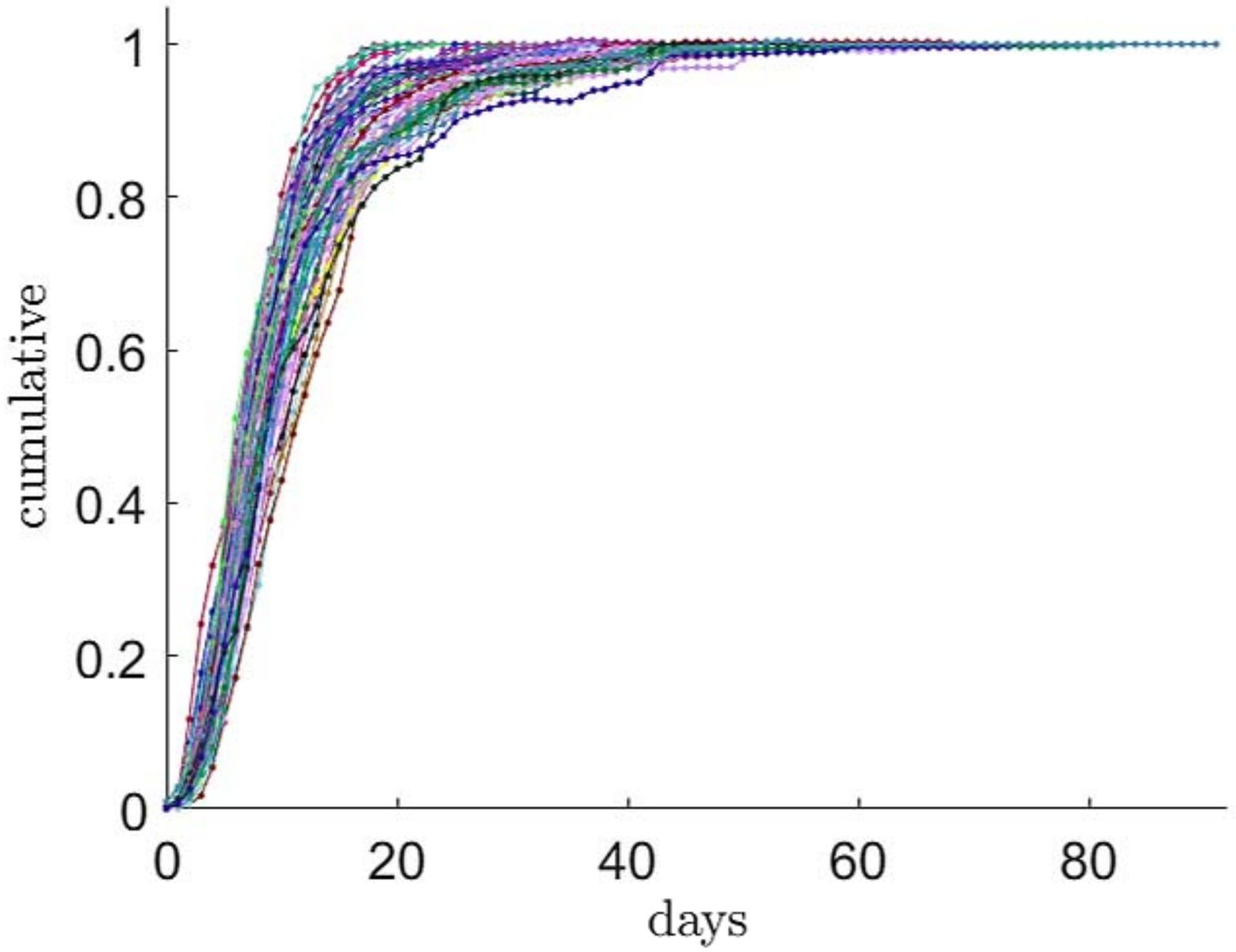} &
    \includegraphics[width=1.125in]{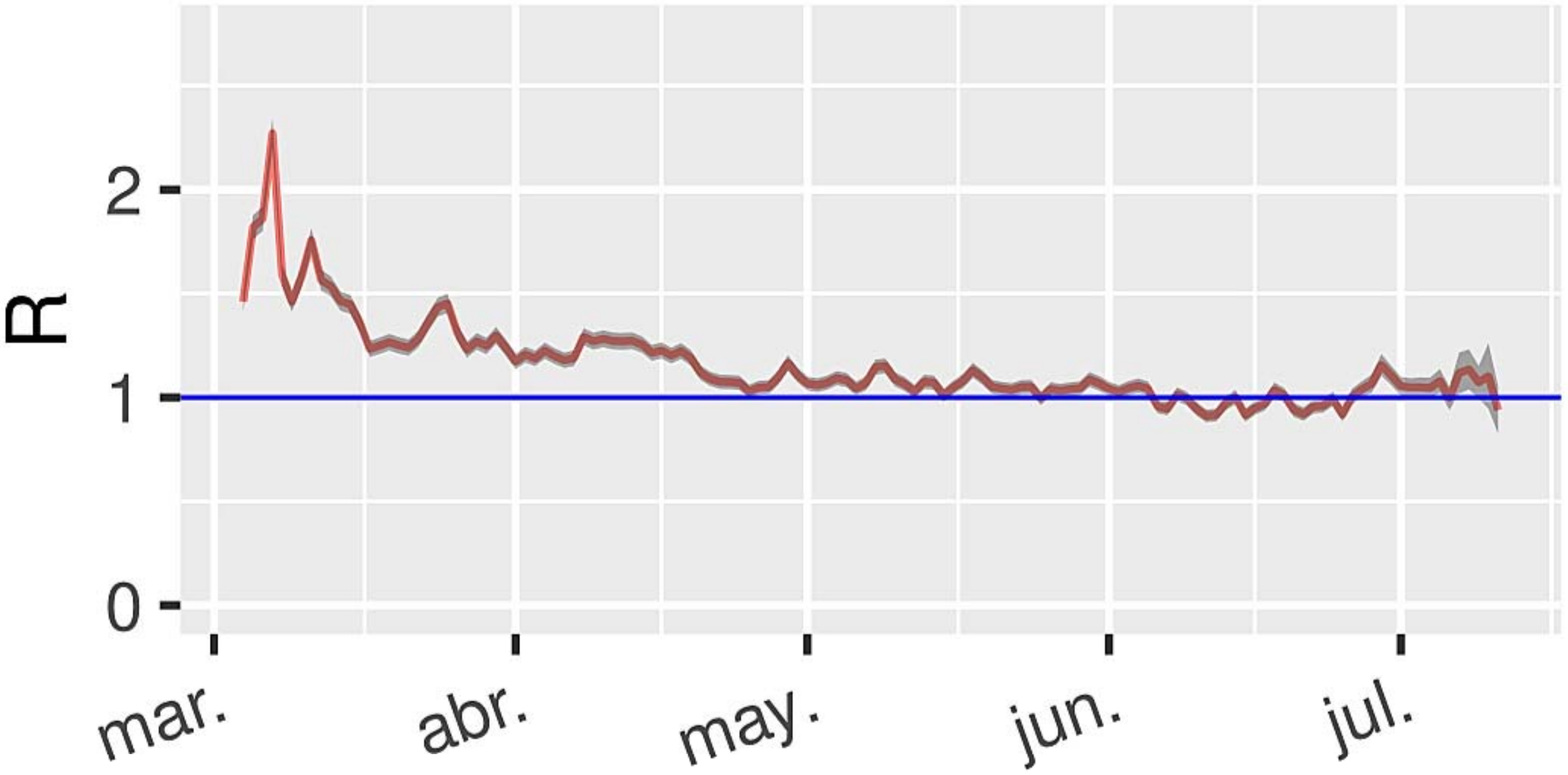} & \includegraphics[width=1.125in]{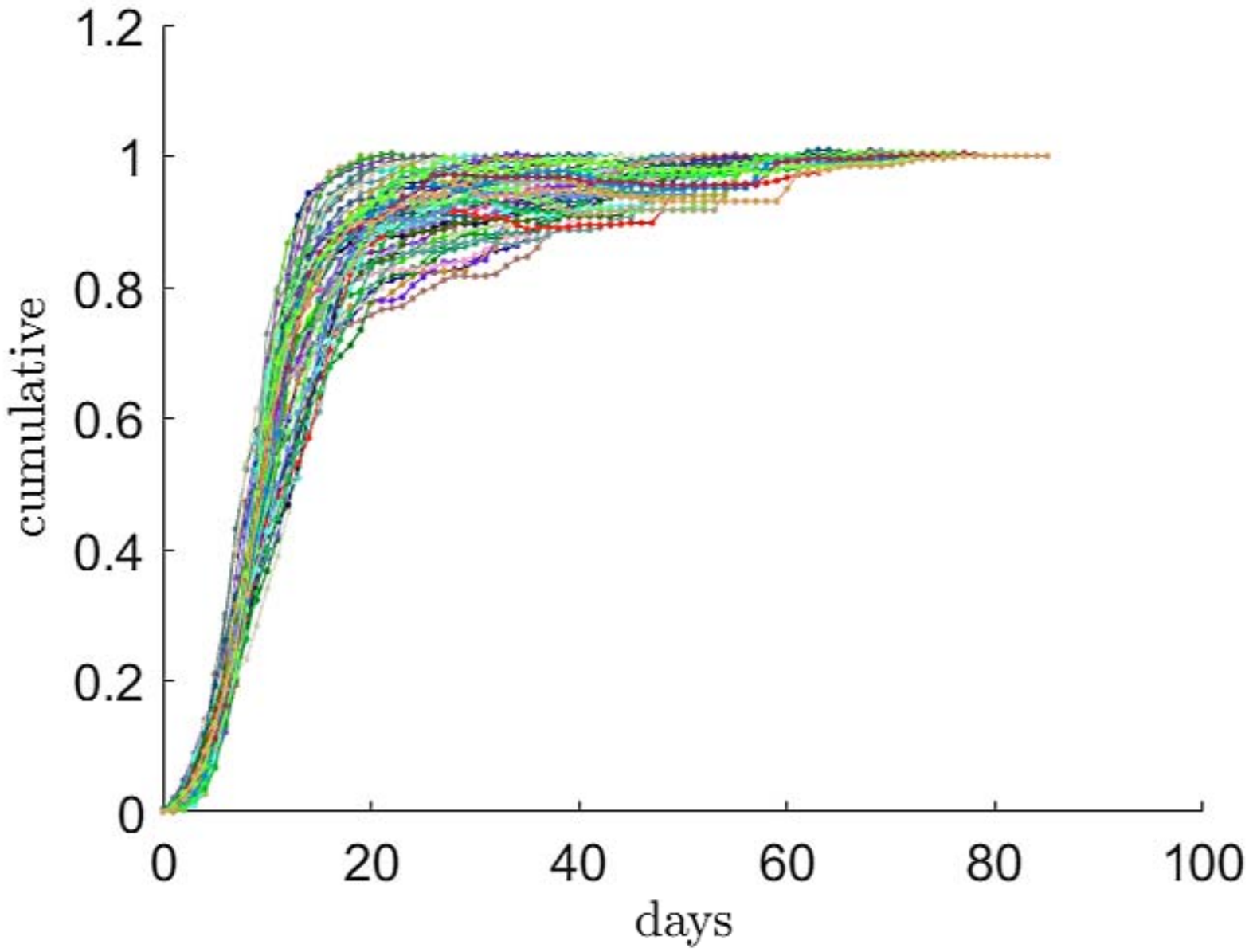}\\
    (a) 1.41 $\pm$ 0.24 & (b) & (c)  0.94 $\pm$ 0.11 & (d)  \\

     \includegraphics[width=1.125in]{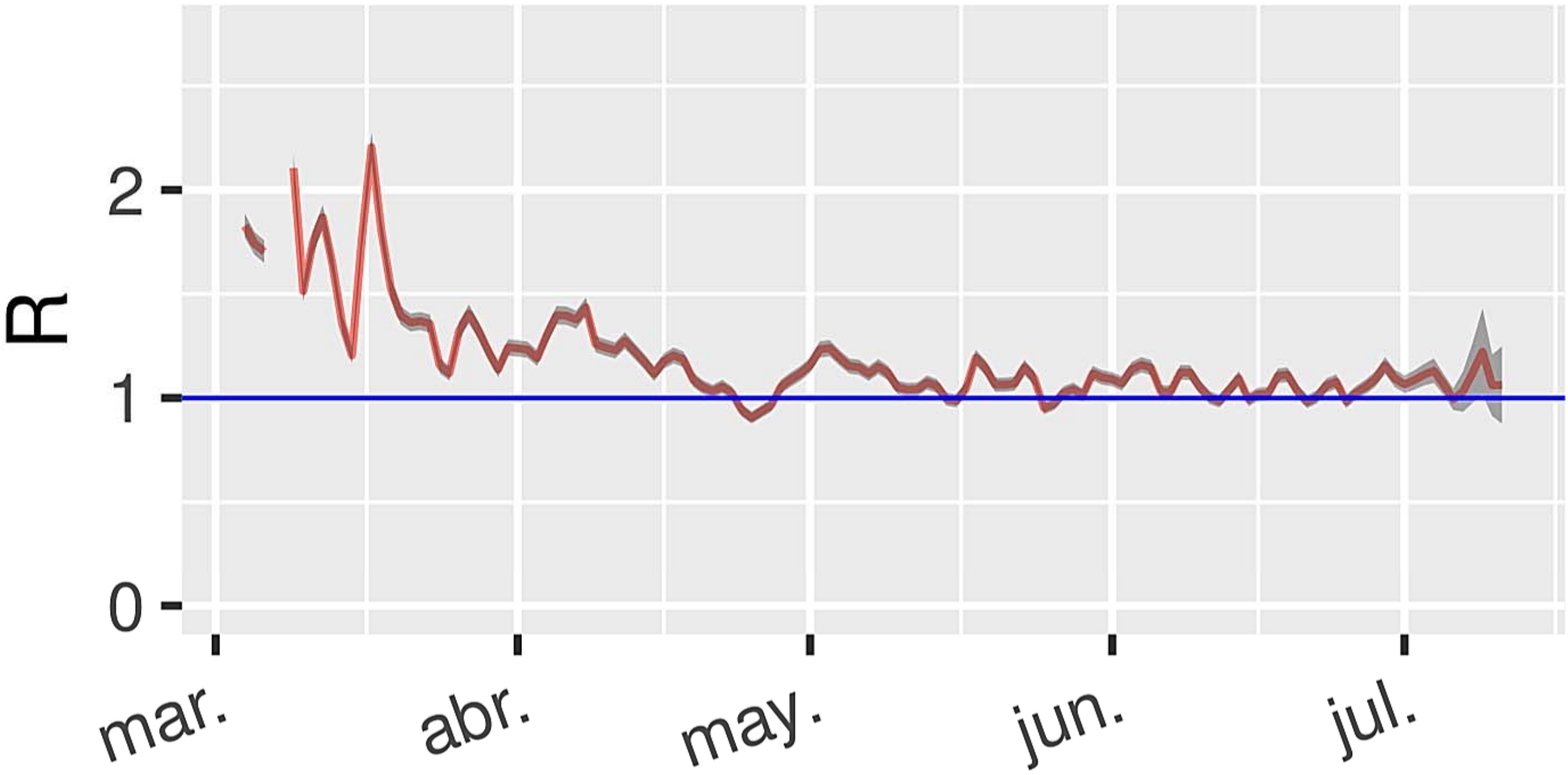} &
    \includegraphics[width=1.125in]{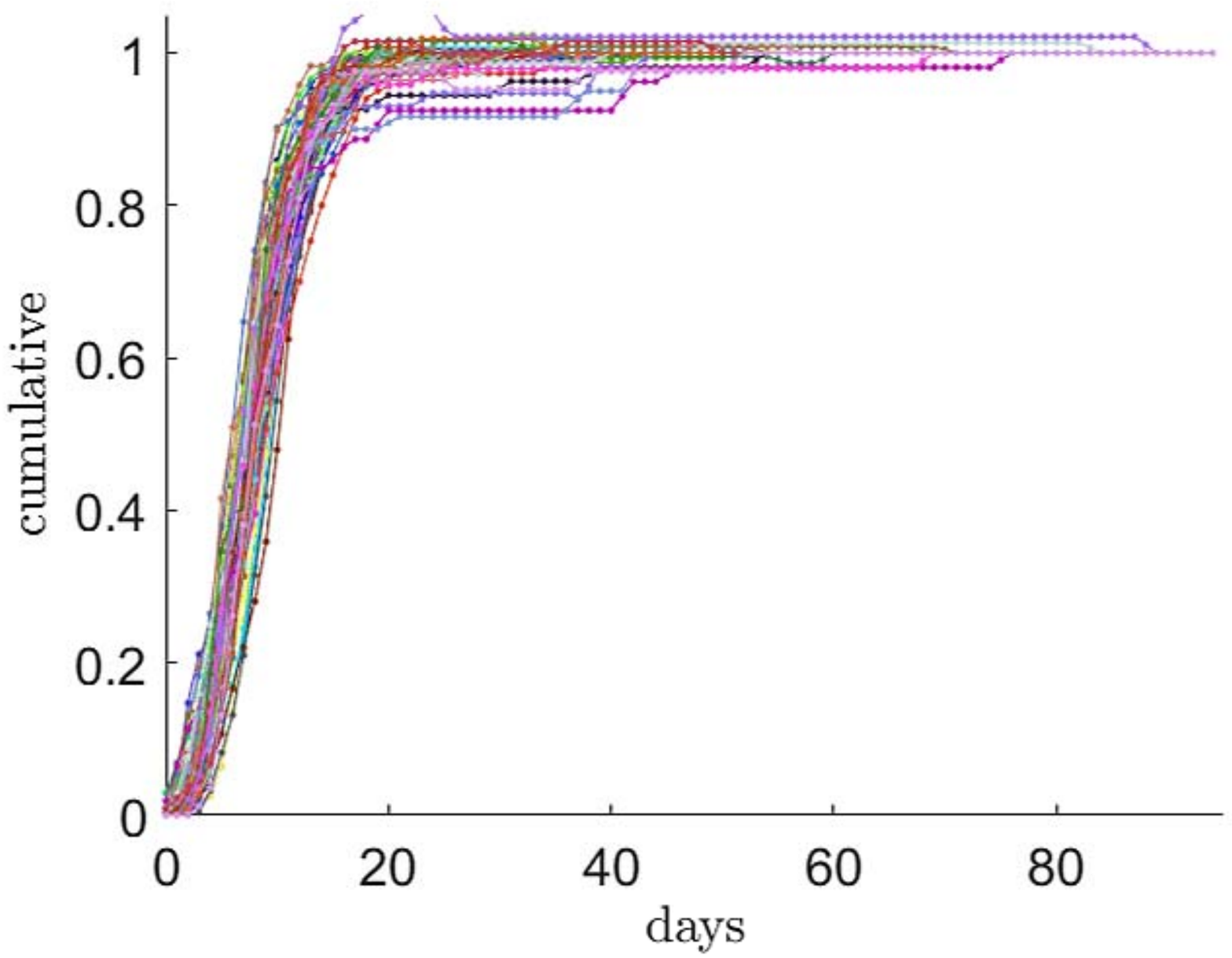} &
    \includegraphics[width=1.125in]{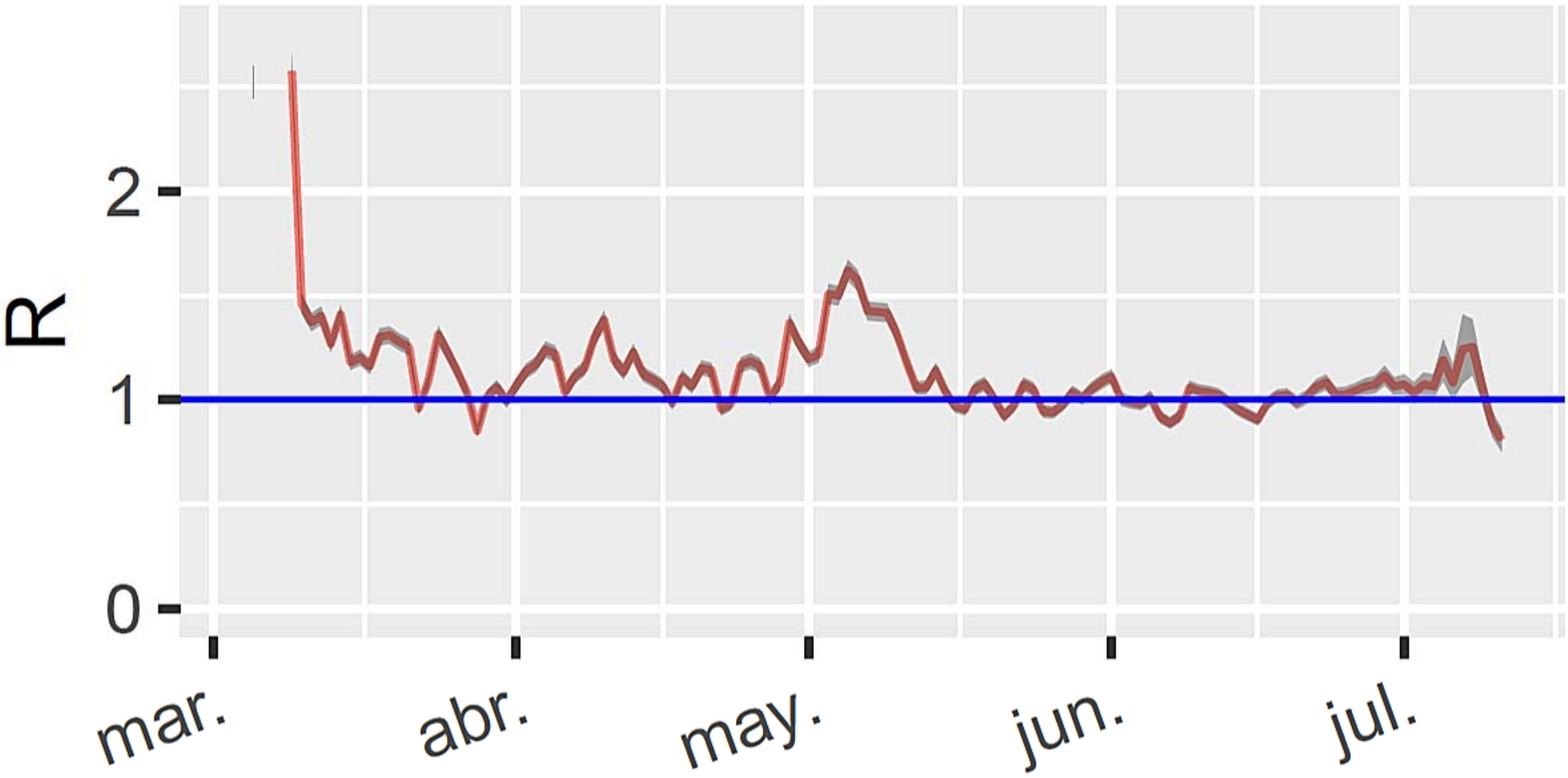} & \includegraphics[width=1.125in]{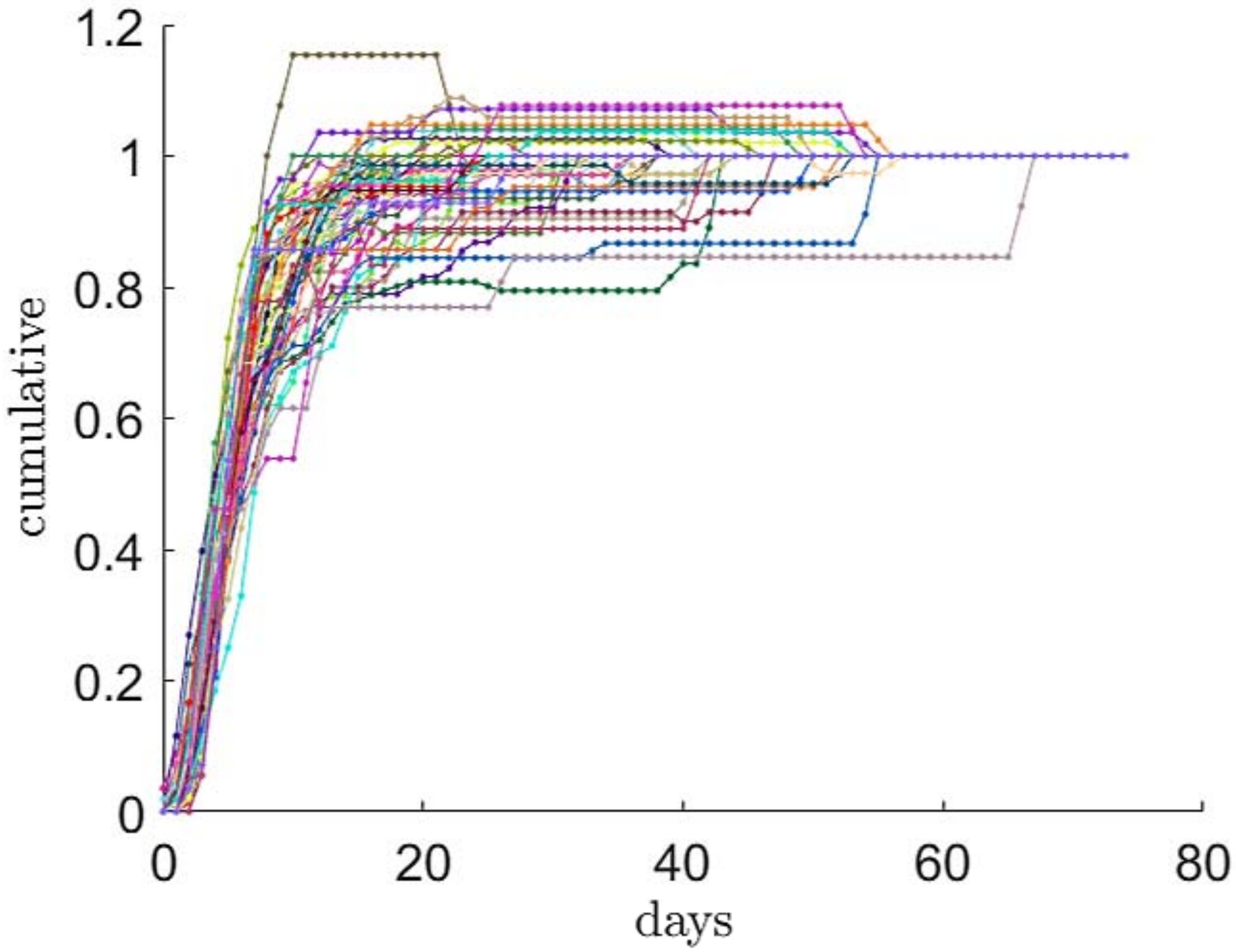}\\
    (e)  1.06 $\pm$ 0.19 & (f)  & (g) 0.81 $\pm$ 0.06 &(h)  \\

    \end{tabular}
    \caption{Estimation of $R_t$ for some states of Mexico and corresponding cumulative normalized distribution for Mexico City (a)-(b), Mexico State (c)-(d), Tabasco (e)-(f), and Queretaro (g)-(h). }
    \label{fig:states}
\end{figure*}

\section{Results}
\label{sec:results}
We took the data set for COVID-19 cases provided by the Mexican Health Ministery corresponding to July 11, 2020. The data set contains 723,668 records, out of which 295,268 correspond to confirmed positives. As time passes by, the number of confirmed positives for a given day $t$ is updated. In Figure~\ref{fig:delay}, we illustrate how each day the updates stack up  a layer of updated registers toward the past. As we accumulate the number of confirmed positive updates, we observe that the total quantity levels off and reaches a maximum at $C_t(D)$ (see Figure~\ref{fig:confirmed-positives}). About 98\% of reports are filled out by day 33. When we divide the daily updates for the day $t$ by $C_t(D)$, we obtain the normalized updates by day and accumulated registers illustrated in Figure~\ref{fig:confirmed-positives}.

We then proceed to  \cred{construct empirical}  distributions \cred{describing}  the variation of $\rho_t^D(\delta)$. We show illustrations of this   for $\delta = 1, 3, 7, 15, 25$ and $35$ in Figure~\ref{fig:fit}. Note that $\delta=0$ is not present as generally the number of reported confirmed positive for $C_t(0) =0$ causing  $\rho_t(0)$ to be undefined.
 Once we have the models for $\rho_t(\delta)$, we may proceed to generate estimates for the number of confirmed positives for $C_t(D)$ using (\ref{eq:confirmed-positives-short}). The mean and standard deviation statistics will provide us with the most likely value and an estimate for the uncertainty. We use the same set of randomly generated values to obtain sequences, which we evaluate using the method proposed by Cori {\it et al.}~\cite{cori2013new} to obtain the instantaneous $R_t$.  Our implementation considers the pre-symptomatic transmission, {\it i.e.}, the incubation period, or the time it takes for an infected person to start showing symptoms, is greater than the latent period, or the time from which an infected person can spread to others. Following Bar-On {\it et al.}~\cite{ bar2020science}, we assume that the latent period lasts for three days and the incubation period for five days.

  We compare the performance of our nowcasting with the proposed by Abbott {\it et al.}~\cite{abbott2020estimating} (see Figure~\ref{fig:comparison}). In their case, the nowcasting tends to closely follow the number of reported confirmed positives, which gives the undesirable effect of resulting in a descending $R_t$, when it is not. Our proposal, on the other hand, increases its certainty naturally as more information is available.

\begin{figure*}
    \centering
    \begin{tabular}{cccc}
    \includegraphics[width=0.925in]{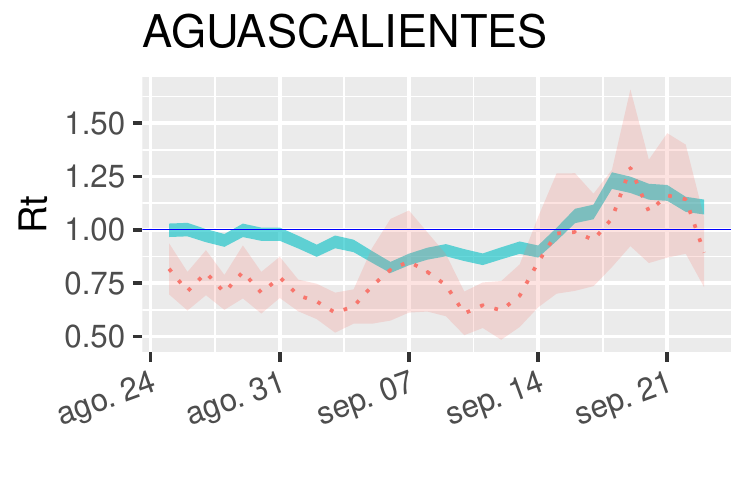} & \includegraphics[width=0.925in]{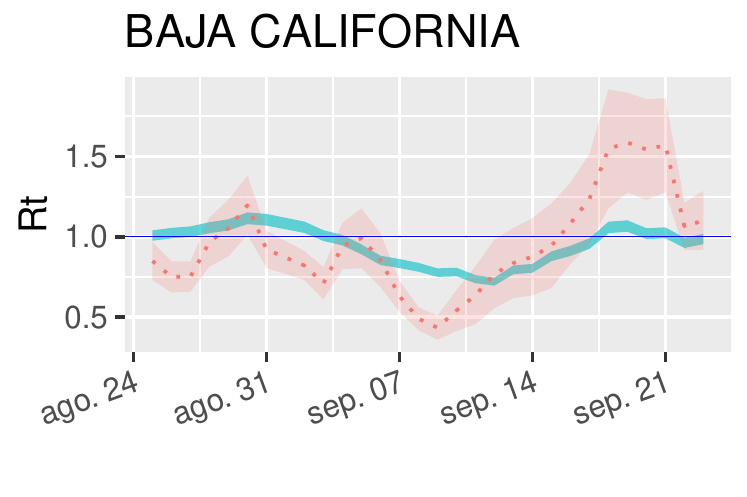} &
    \includegraphics[width=0.925in]{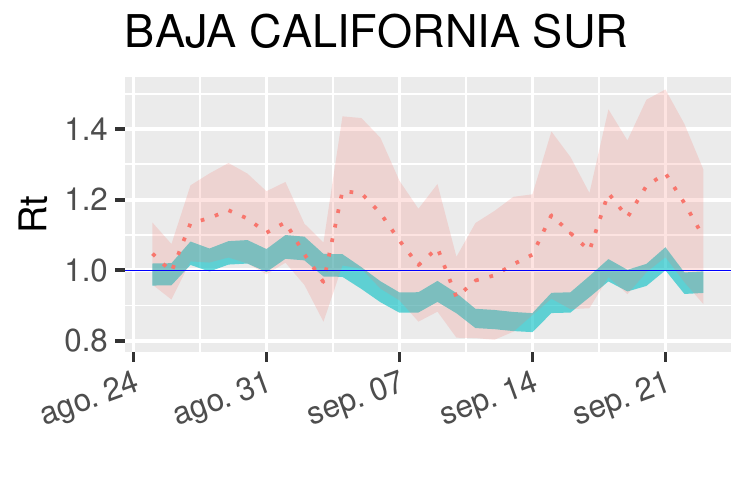} & 
    \includegraphics[width=0.925in]{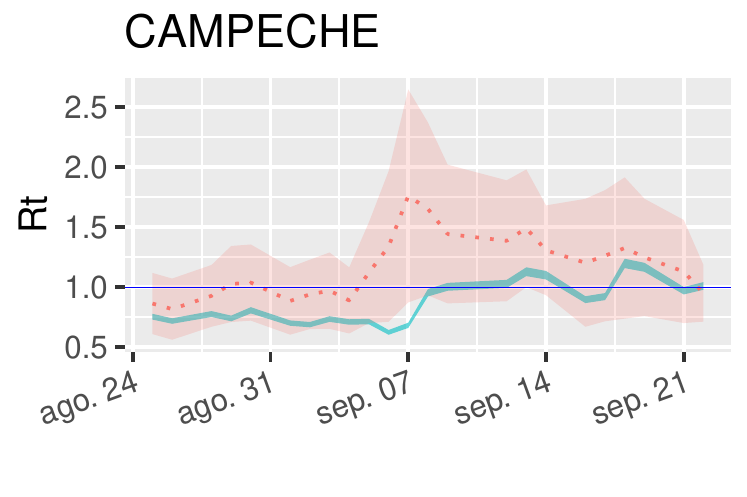}\\
    \begin{scriptsize}
    (a)
    \end{scriptsize}& 
    \begin{scriptsize}(b)
    \end{scriptsize}& 
    \begin{scriptsize}(c)
    \end{scriptsize}& 
    \begin{scriptsize}(d)
    \end{scriptsize}\\
    \includegraphics[width=0.925in]{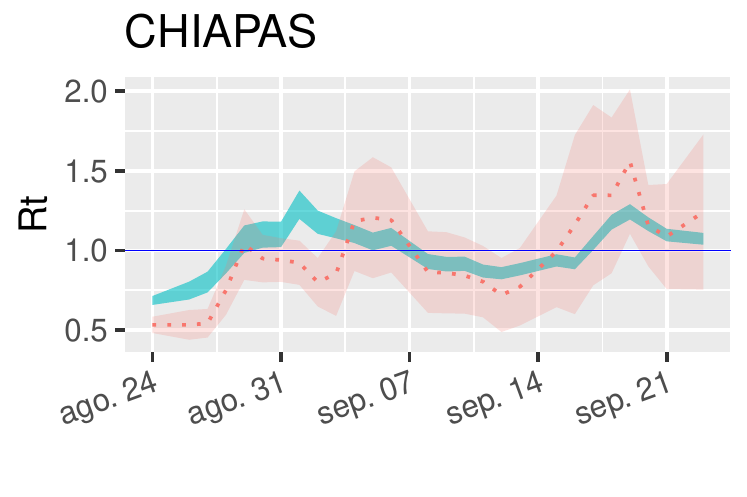} & 
    \includegraphics[width=0.925in]{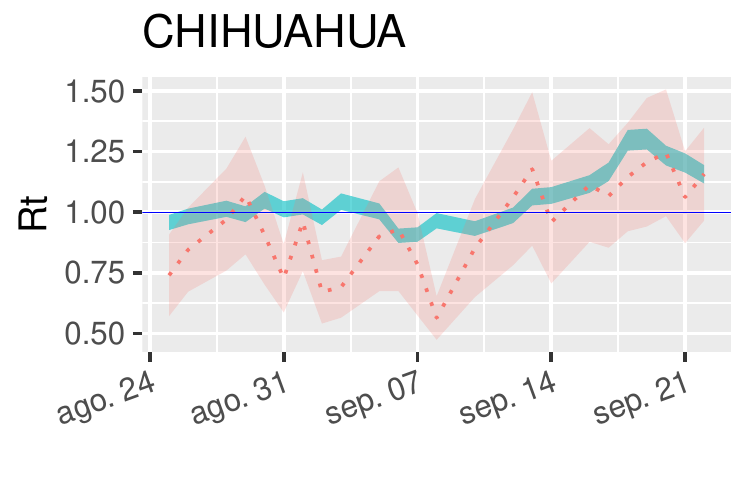} &
    \includegraphics[width=0.925in]{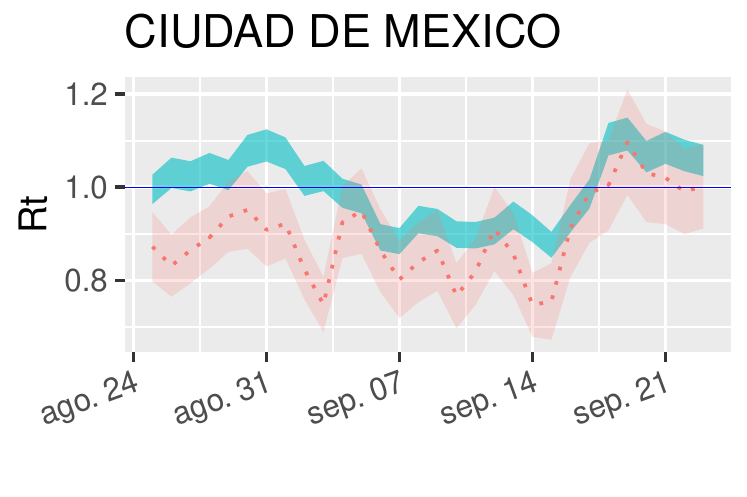} & \includegraphics[width=0.925in]{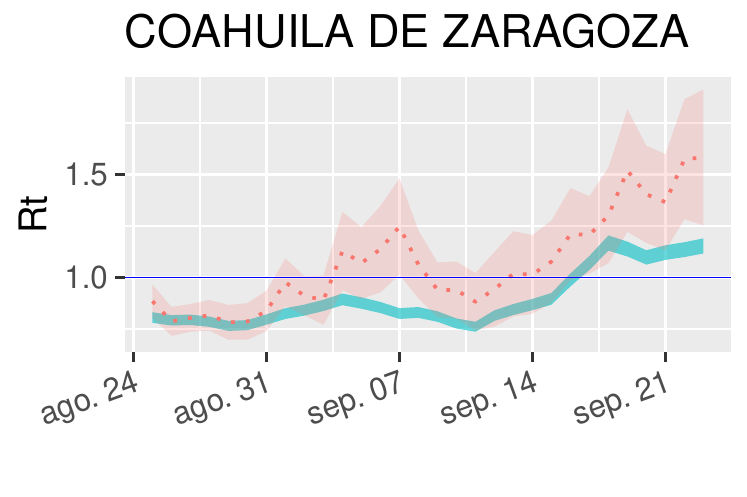} 
     \\
    
    \begin{scriptsize}
    (e)
    \end{scriptsize}& 
    \begin{scriptsize}(f)
    \end{scriptsize}& 
    \begin{scriptsize}(g)
    \end{scriptsize}& 
    \begin{scriptsize}(h)
    \end{scriptsize}\\
     \includegraphics[width=0.925in]{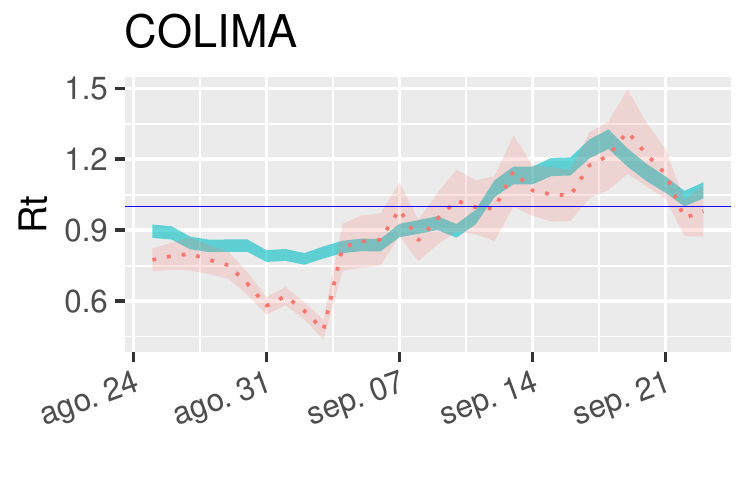}    & \includegraphics[width=0.925in]{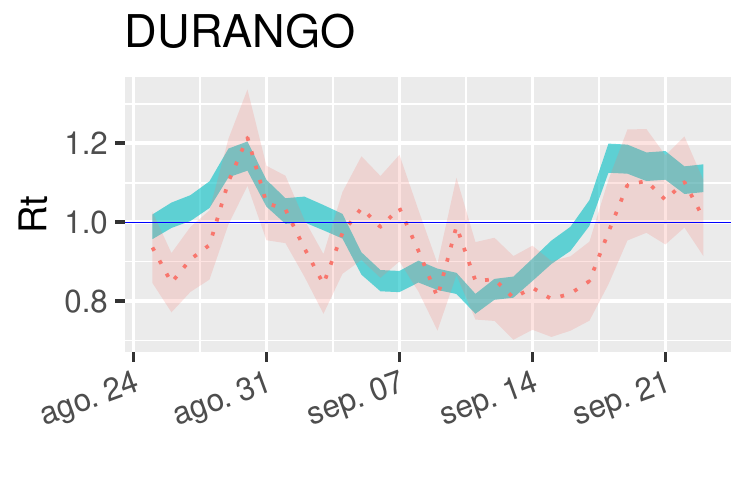} &
    
    \includegraphics[width=0.925in]{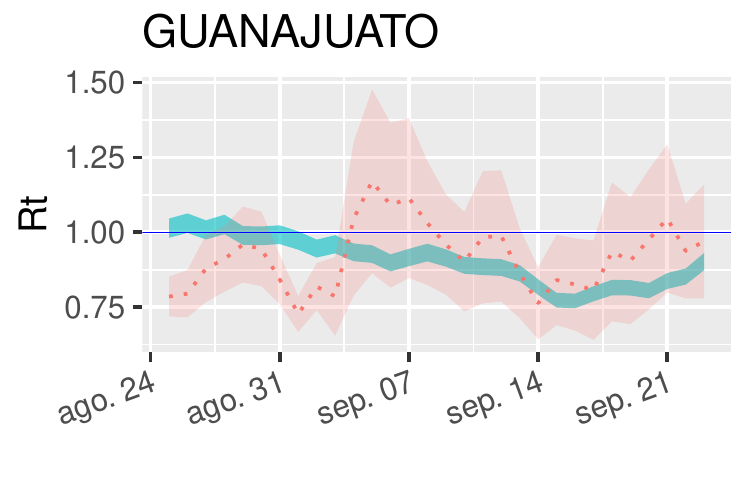}& 
    \includegraphics[width=0.925in]{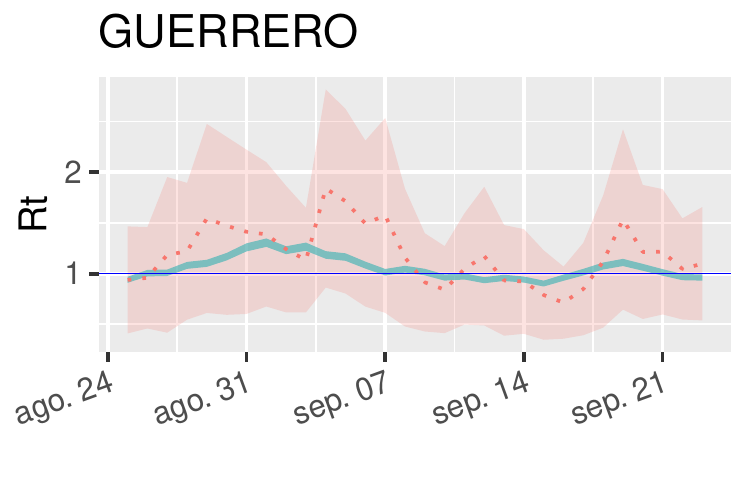}\\
    \begin{scriptsize}(i)
    \end{scriptsize}& 
    \begin{scriptsize}(j)
    \end{scriptsize}& 
    \begin{scriptsize}(k)
    \end{scriptsize}& 
    \begin{scriptsize}(l)
    \end{scriptsize}\\
    \includegraphics[width=0.925in]{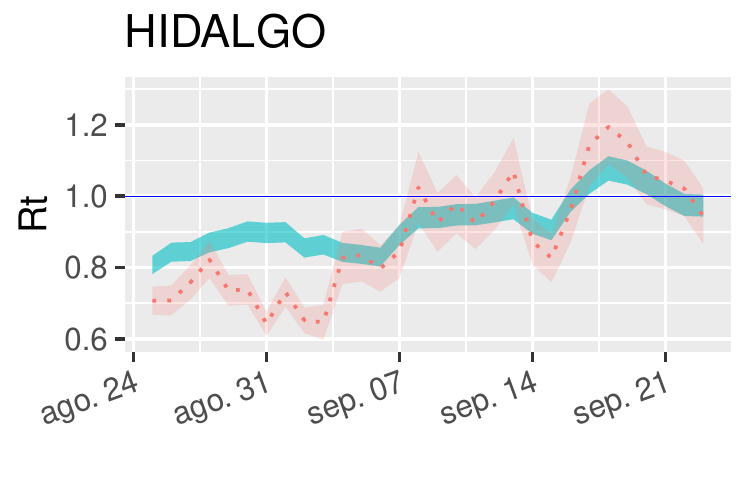} & \includegraphics[width=0.925in]{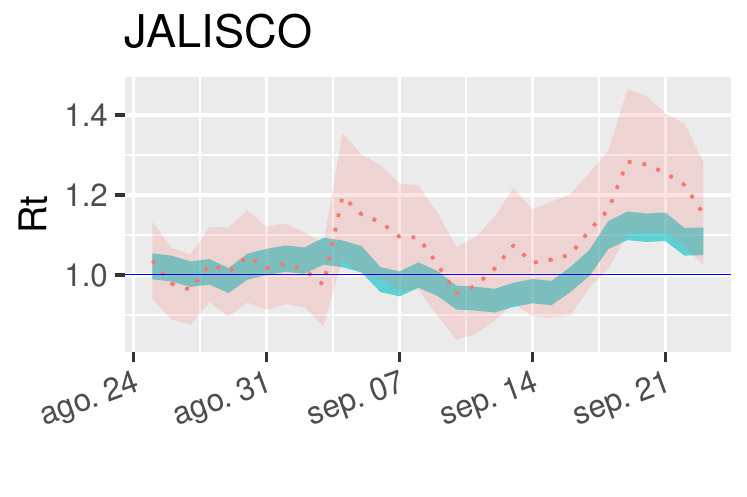} &
    \includegraphics[width=0.925in]{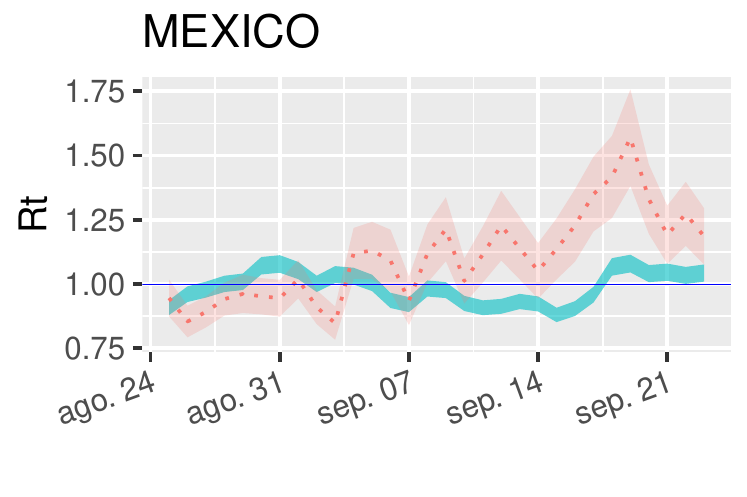} & 
    \includegraphics[width=0.925in]{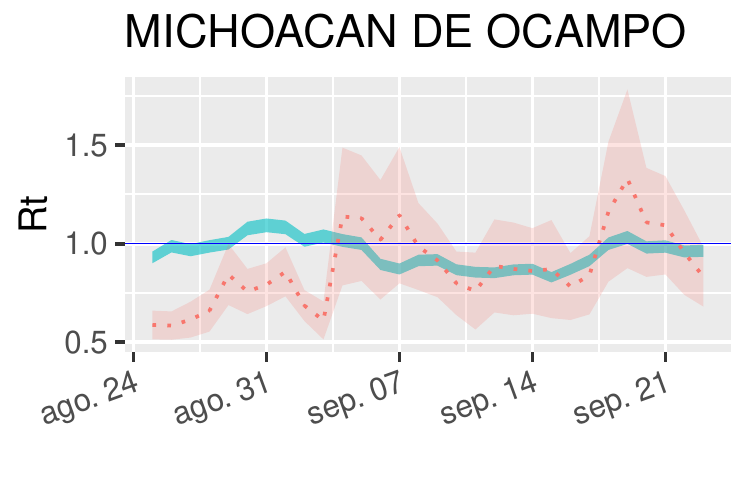}\\
    \begin{scriptsize}(m)
    \end{scriptsize}& 
    \begin{scriptsize}(n)
    \end{scriptsize}& 
    \begin{scriptsize}(o)
    \end{scriptsize}& 
    \begin{scriptsize}(p)
    \end{scriptsize}\\
    \includegraphics[width=0.925in]{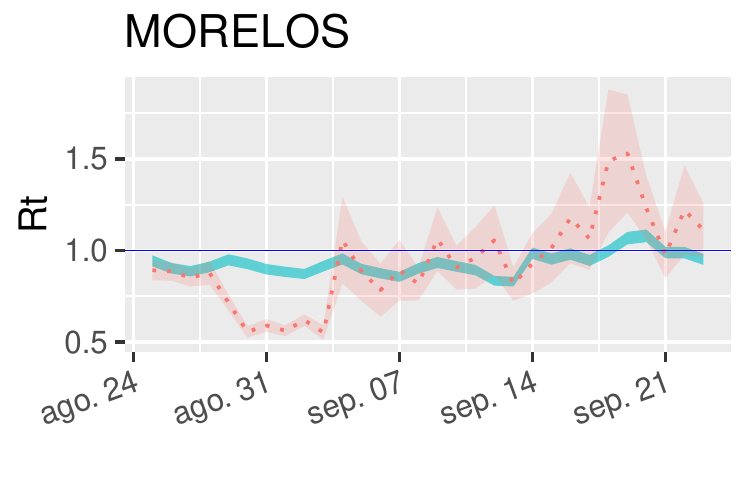} & 
    \includegraphics[width=0.925in]{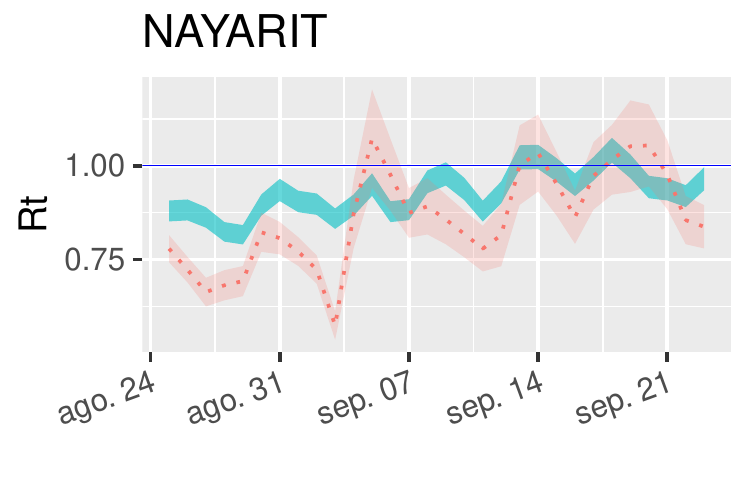}&
    \includegraphics[width=0.925in]{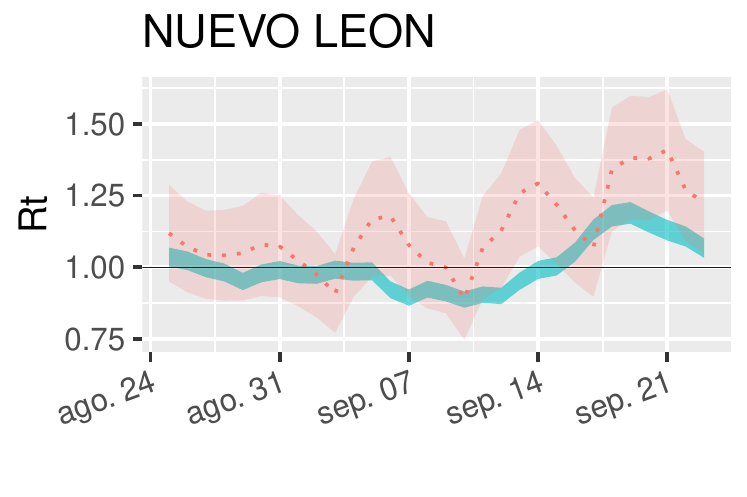} & 
    \includegraphics[width=0.925in]{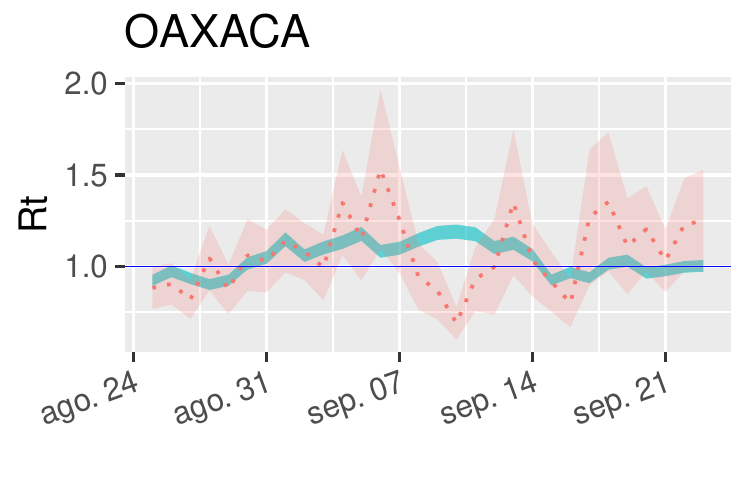}\\
    \begin{scriptsize}(q)
    \end{scriptsize}& 
    \begin{scriptsize}(r)
    \end{scriptsize}& 
    \begin{scriptsize}(s)
    \end{scriptsize}& 
    \begin{scriptsize}(t)
    \end{scriptsize}\\
    \includegraphics[width=0.925in]{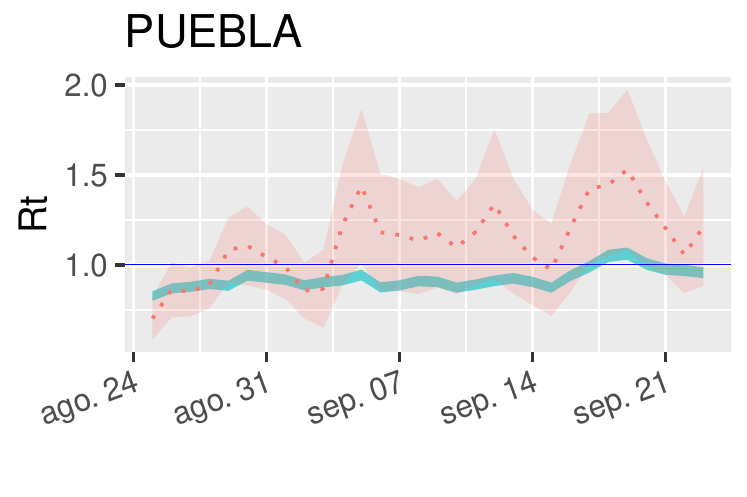} & \includegraphics[width=0.925in]{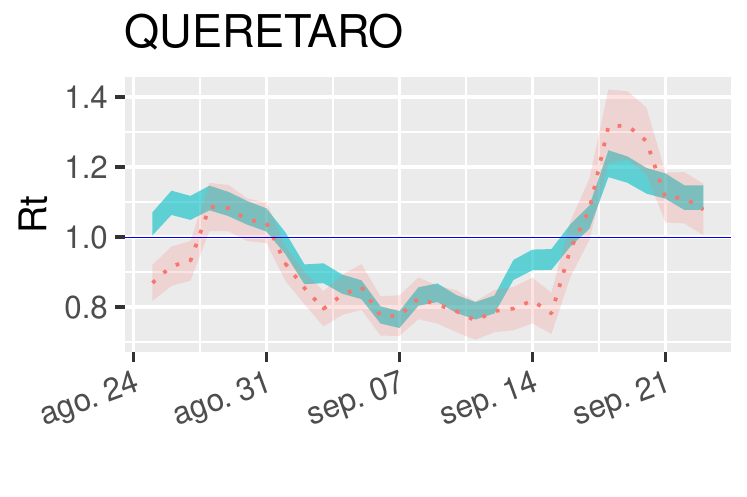}
    &
    \includegraphics[width=0.925in]{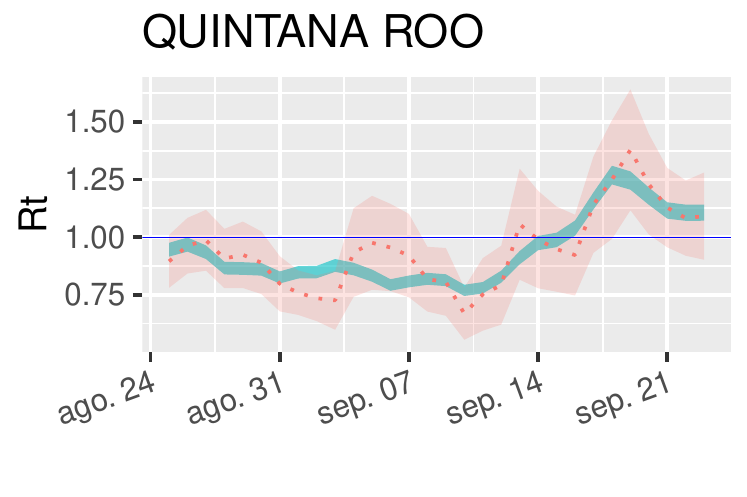}
     & 
      \includegraphics[width=0.925in]{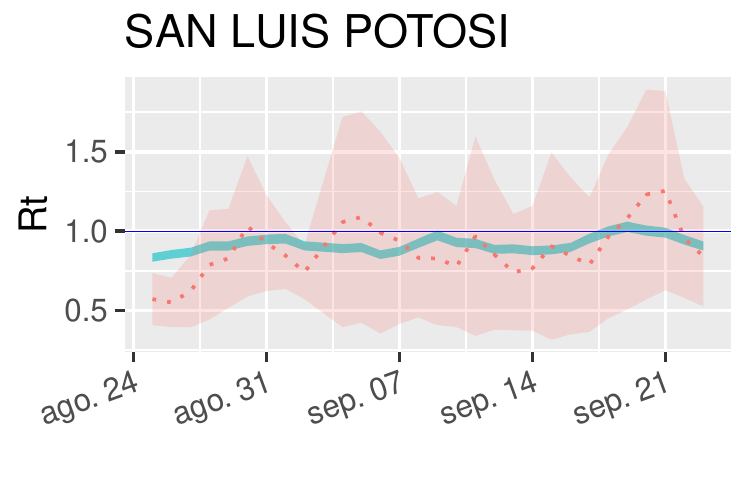}
    \\
    \begin{scriptsize}(u)
    \end{scriptsize}& 
    \begin{scriptsize}(v)
    \end{scriptsize}& 
    \begin{scriptsize}(w)
    \end{scriptsize}& 
    \begin{scriptsize}(x)
    \end{scriptsize}\\
    \includegraphics[width=0.925in]{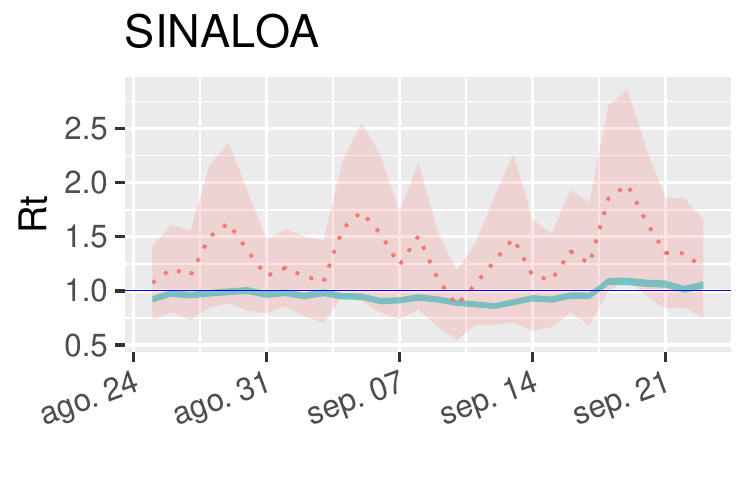}& \includegraphics[width=0.925in]{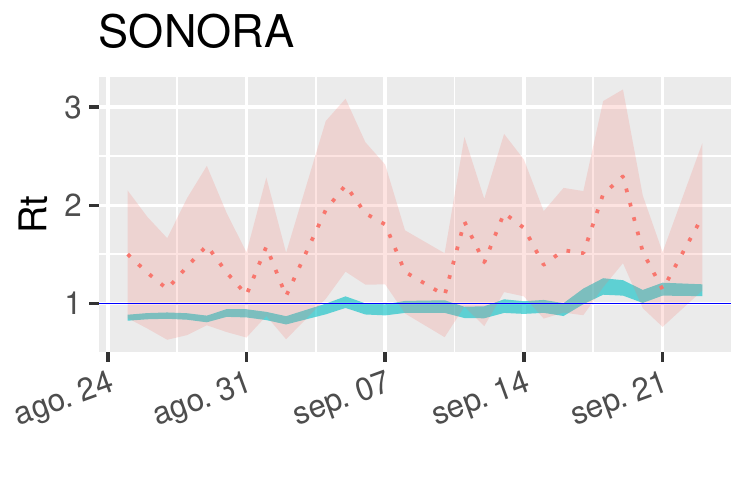} &
    \includegraphics[width=0.925in]{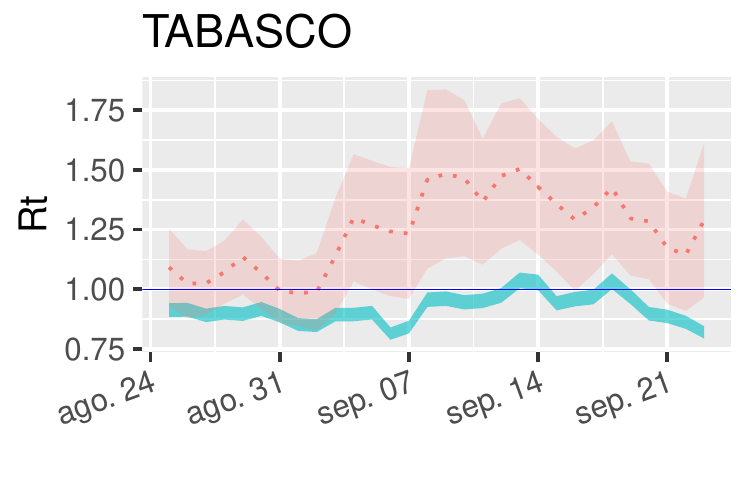} &
    \includegraphics[width=0.925in]{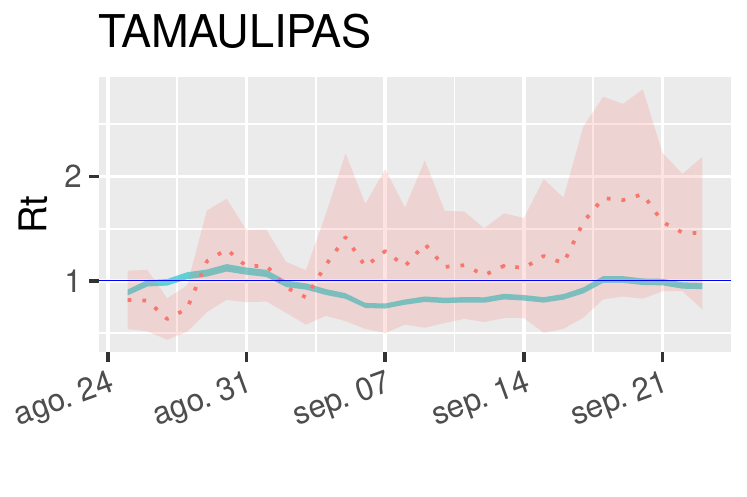}\\
    
    \begin{scriptsize}(y)
    \end{scriptsize}& 
    \begin{scriptsize}(z)
    \end{scriptsize}& 
    \begin{scriptsize}($\alpha$)
    \end{scriptsize}& 
    \begin{scriptsize}($\beta$)
    \end{scriptsize}\\
       \includegraphics[width=0.925in]{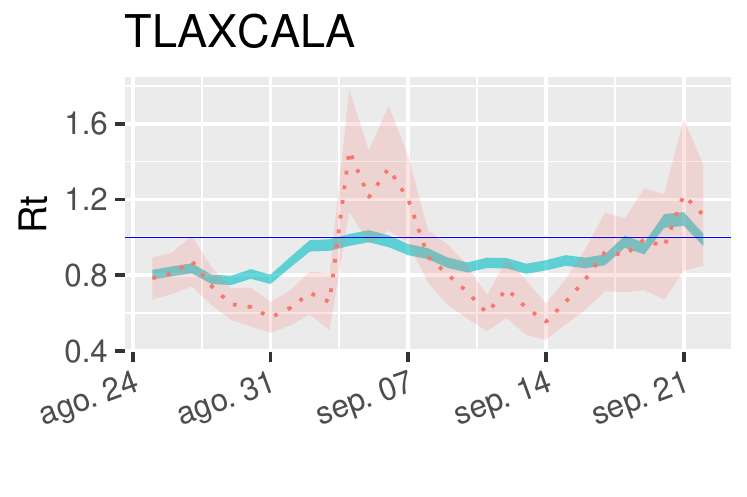}  & \includegraphics[width=0.925in]{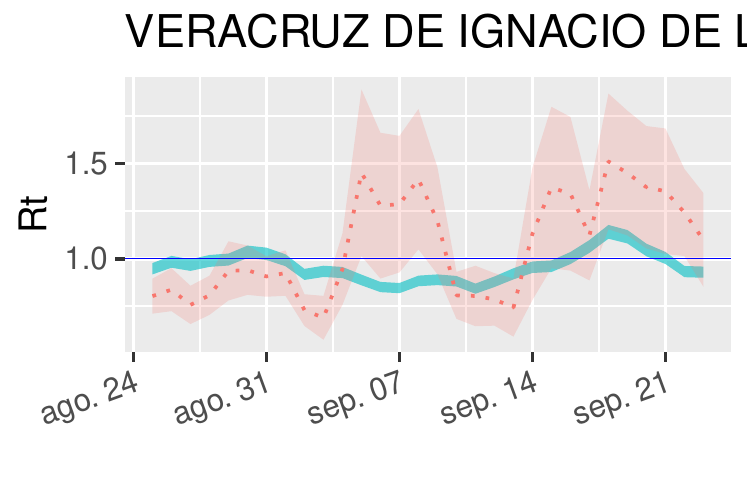} &
    \includegraphics[width=0.925in]{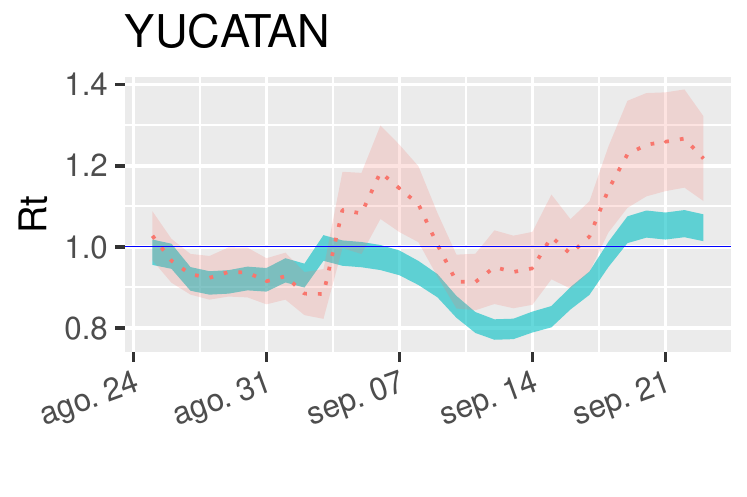} & 
    \includegraphics[width=0.925in]{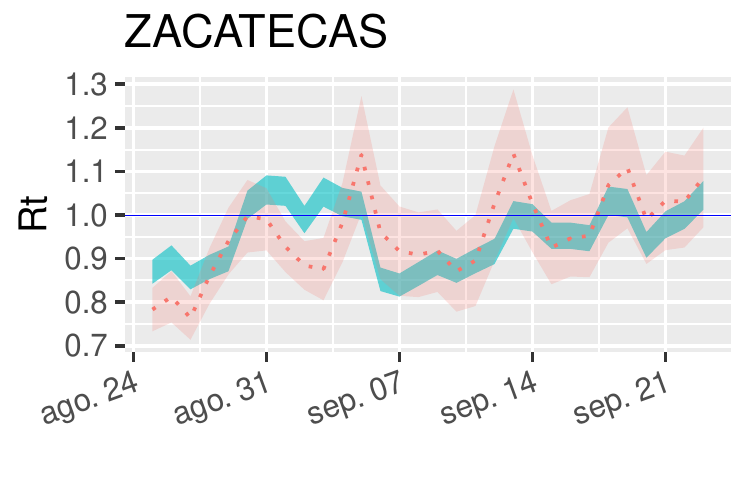}\\
    \begin{scriptsize}($\gamma$)
    \end{scriptsize}& 
    \begin{scriptsize}($\delta$)
    \end{scriptsize}& 
    \begin{scriptsize}($\epsilon$)
    \end{scriptsize}& 
    \begin{scriptsize}
    ($\zeta$) 
    \end{scriptsize}\\

    \end{tabular}
    \caption{Effective reproduction number $R_t$. On November 26, 2020, we revisited the $R_t$ (green band) that occurred and the $\hat{R_t}$ predicted (red band) between August 24 and September 23. The green band is thick because there were still some updates, affecting registers three months before.}
    \label{fig:performance}
\end{figure*}

\begin{figure*}
    \centering
    \begin{tabular}{cc}
    \includegraphics[width=2.25in]{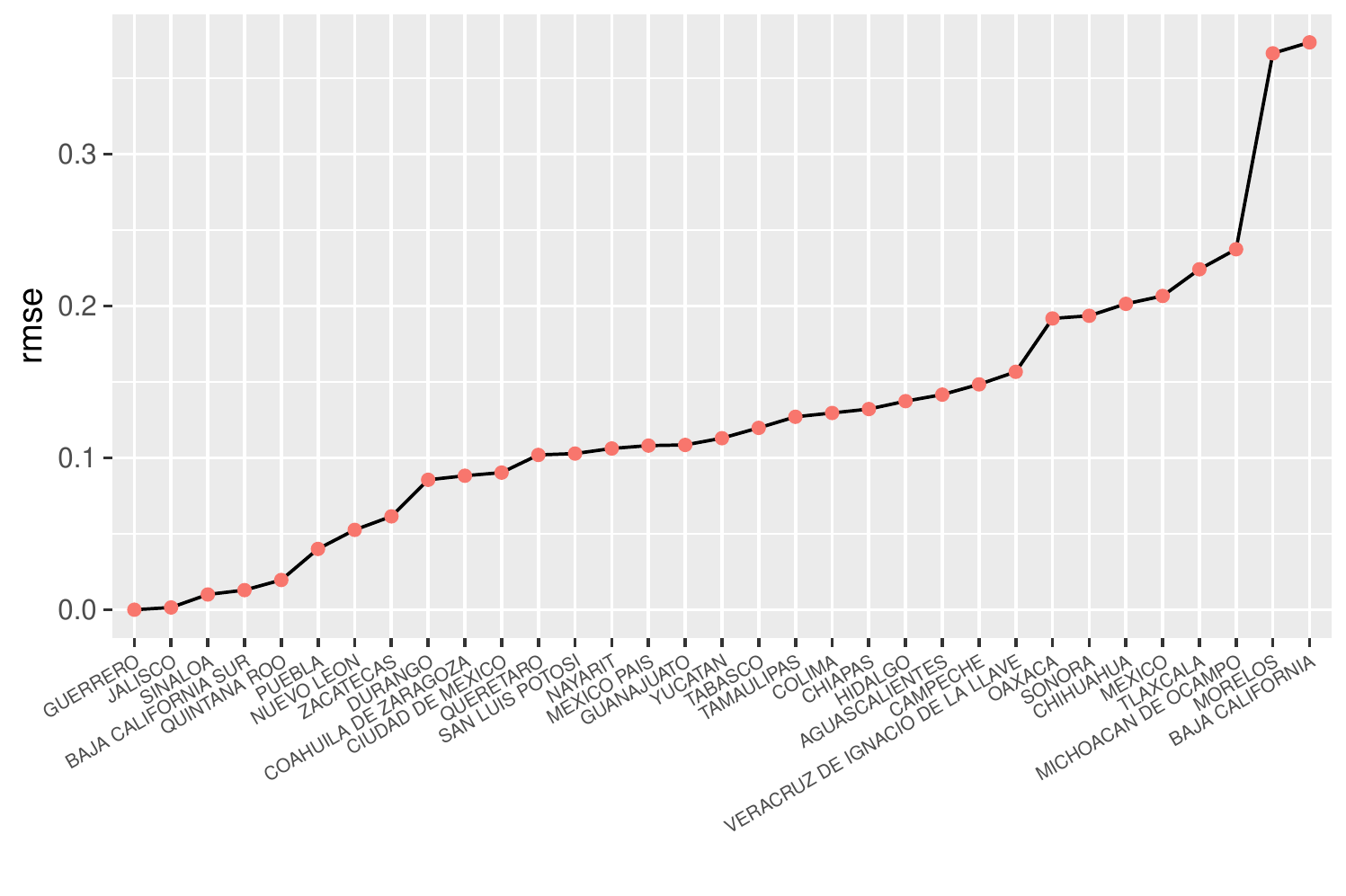} &
    \includegraphics[width=2.25in]{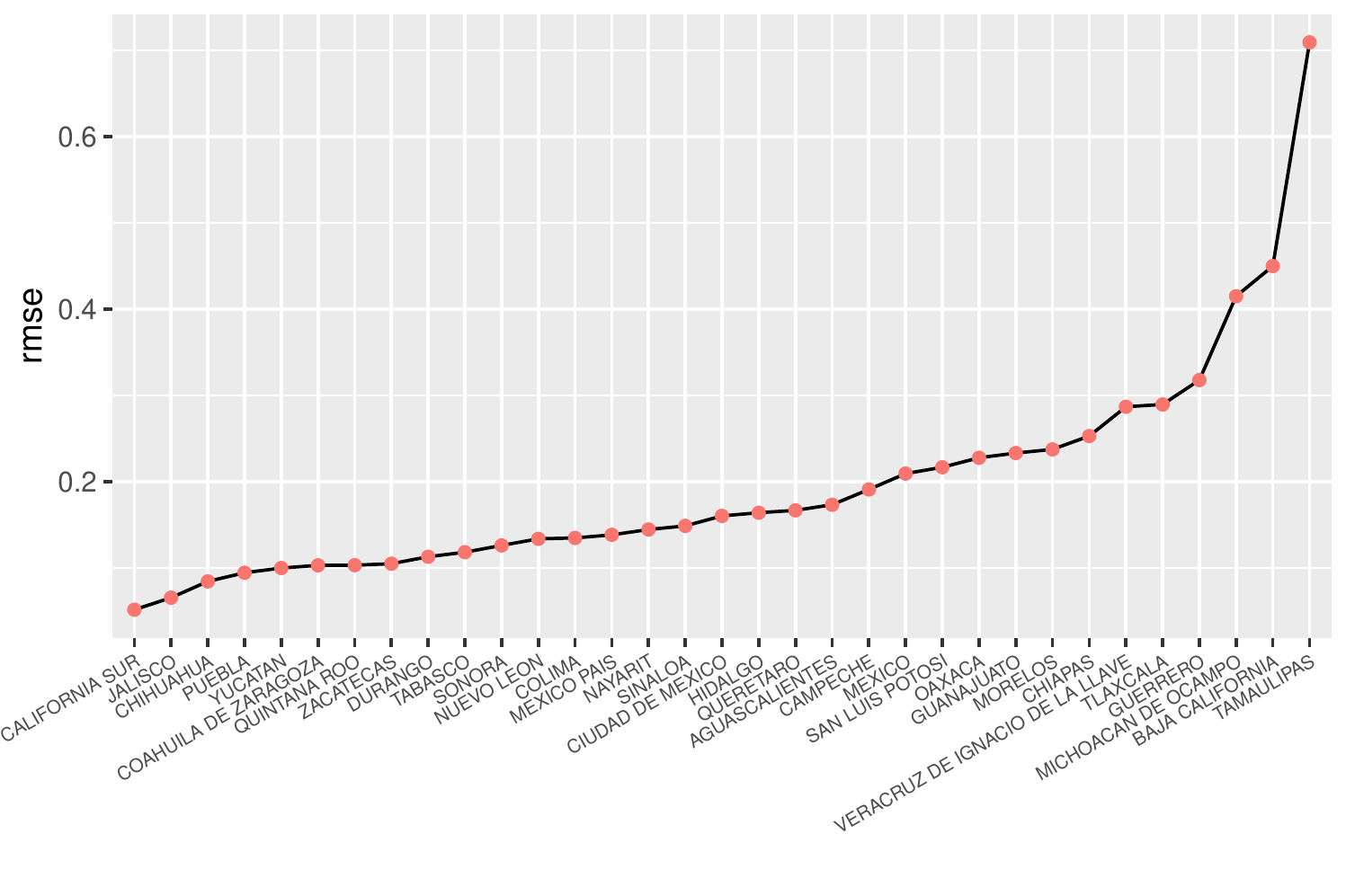}\\
    (a) Full dataset & (b) Last 20 days\\
    \end{tabular}
    \caption{ \cred{Nowcasting performance. We computed the RMSE for the values out of the band defined between the mean $\overline{R}_t$ and one standard deviation $\sigma$. In (a), we use all the distributions of cases available for nowcasting. In (b), we use the most recent 20 distributions. }}
    \label{fig:rmse}
\end{figure*}

As we increase $\delta$ in our nowcasting exercise, there is a tendency to observe fewer cases, because there has not been enough time for the information to arrive. We set a dynamic threshold to stop the nowcasting estimation  when for a particular day of analysis, $\delta$, the number of confirmed positive cases is less than 30. Also, we have observed that as the number of confirmed positive is less,  the normalized cumulative curves tend to be noisier. In Figure~\ref{fig:states}, we illustrate what happens for entities in Mexico where the number of confirmed positive cases is 59,667 (Mexico City), 44,114 (State of Mexico),  15,909 (Tabasco), and 2,667 (Querétaro). We believe that our method works best when the number of positive cases is beyond 2,600 for the observed interval of 90 days. Using this threshold, there are still currently 30 States (out of 32) and 32 Municipios (out of 2450) in Mexico subject to our analysis.

\cred{To assess our scheme's performance quantitatively, we analyzed data in the past, when the uncertainty in the estimation of the effective reproduction number, $R_t$, is small, and compare it with our predictions at that date. On November 25, 2020, we observed a period starting three months before, from August 24 to September 23, and compared our nowcast prediction $\hat{R}_t$  with $R_t$ for each of the 32 states of Mexico (see Figure~\ref{fig:performance}). Our method outputs $\hat{R}_t$ as a distribution which spread grows as the prediction approaches the current date. For the performance assessment, we characterize the distribution of  $\hat{R}_t$ with its mean $\overline{R}_t$ and one standard deviation at each side. To evaluate the performance,  we obtain the root mean squared error, RMSE, between  $R_t$ and the prediction band created by $\overline{R}_t$ and one standard deviation $\sigma$ (see Figure~\ref{fig:rmse}(a)) as
\begin{equation}
\mbox{RMSE} = \sqrt{\frac{1}{n}\left(
\sum_{R_t^i  <  \overline{R}_t^i - \sigma_i} (R_t^i  - (\overline{R}_t^i - \sigma_i))^2 + \sum_{R_t^i  >  \overline{R}_t^i + \sigma_i} (R_t^i  - (\overline{R}_t^i + \sigma_i))^2\right)},     
\end{equation}
\cred{where $n$ is the number of points that meet the logical conditions.} 
One observes that for states such as Guerrero, Jalisco, and Sinaloa, with an RMSE of 0.0, 0.001, 0.010, the band of uncertainty frequently includes the value of $R_t$, while for Michoacan de Ocampo, Morelos and Baja California, with an RMSE of 0.237, 0.366, 0.373,  the value of $R_t$ is sometimes outside the band of uncertainty.  
}

\cred{An intriguing question is whether one should employ the whole sequence of reports to construct the frequency distributions for $\rho$ or select the more recent ones, under the rationale that the infectious dynamics have changed or health institutions have implemented new reporting practices. To study this effect, we repeated the performance evaluation previously described but used the last 20 available distributions. In Figure~\ref{fig:rmse}(b), with a maximum RMSE above 0.7 and a generally more step curve, we show that the performance declines when we use the last observations compared with using the full set.   }

\section*{\cred{Discussion and }Conclusion}
\cred{Lack of testing is a significant issue in Mexico. Despite frequent suggestions by the World Health Organization, the number of tests performed normalized by its populations is low among the worst-hit countries~\cite{shams2020analyzing}. Thus, a data-driven approach, such as ours, is likely to underrepresent the phenomenon's true nature. Also, we need further studies to assess the effects of novel testing methods with potentially faster turnaround and the implementation of improved procedures to generate, process, analyze, and transfer information. However, our evidence suggests that our method works best, using even the information developed since the epidemy's onset.   }

In this document, we have presented a nowcasting method to estimate the number of confirmed positives. We have shown that this may be the foundation to generate plausible sequences out of which one may  determine useful epidemy tracking indicators, such as the basic reproduction number. Our method naturally expresses uncertainty due to the lack of information but eventually gains certainty as more data accumulates.

Our method's strength is that it is based on the self-reported onset of symptoms, in contrast to other methods that use the number of confirmed positives cases accumulated by the report's day to infer this quantity. A potential drawback of our approach is that it relies on a regularity of the update cycle. As researchers implement more sophisticated systems for testing and reporting, the statistics may change. To remedy this potential effect, one may eliminate old observations and update the distributions for $\rho_t$ regularly. Due to the difference between the incubation and latent periods, and delays in the detection and reporting cycle,   our model estimates $R_t$ up to several days in the past. We decided to take no further assumptions about the progression of the epidemy. Although potentially some form of state estimation may be possible to implement to fill the gap.

We believe that it is crucial to continue developing solutions to quickly, robustly, and reliably estimate indicators such as the basic reproduction number. A possible direction for future research may be to determine the disaggregation level to continue to generate a reliable indicator. The resulting nowcasting methods should compensate for the delays inherent in producing and processing information about this critical, global, and urgent problem.
Also, we are planning to study the extend at which our model can be incorporated into dynamics-based models. This enhancement could offer improved nowcasting.

\begin{acks}
We thank Carlo Tomasi, Duke University, for providing the fundamental idea for our approach. Thanks to Dagoberto Pulido for implementing Abbott {\it et al.}~\cite{abbott2020estimating} to generate (c).  
\end{acks}

\begin{biog}
Joaquín Salas is a professor in the field of Computer Vision at Instituto Politécnico Nacional. Member of the Mexican National Research System, his research interests include the monitoring of natural systems using visual perception and aerial platforms. Salas received a Ph.D. in computer science from ITESM, México. He has been a visiting scholar at Stanford University, Duke University, Oregon State University, Xerox PARC, the Computer Vision Center, and the École Nationale Supérieure des Télécommunications de Bretagne. He has served as co-chairperson of the Mexican Conference for Patter Recognition three times.  Salas was Fulbright scholar for the US State Department. He has been invited editor for Elsevier Pattern Recognition and Pattern Recognition Letters. For his services at the Instituto Politécnico Nacional, he received the {\it Lázaro Cárdenas} medal  from the President  of Mexico.
\end{biog}

\begin{dci}
The author declared no potential conflicts of interest with respect to the research, authorship, and/or publication of this
article.
\end{dci}

\begin{funding}
 This work was partially
funded by SIP-IPN 20201357.
\end{funding}

\begin{sm}
 To foster further research, allowing other researchers to verify our results and serve as a stepping stone, we make our code publicly available at \url{https://www.github.com/joaquinsalas/nowcastingRt}.
\end{sm}

\bibliographystyle{SageV}

\bibliography{references}

\end{document}